\RequirePackage{fix-cm}

\documentclass[twocolumn]{svjour3}          
\smartqed  
\usepackage{url}
\usepackage{caption}
\usepackage{subcaption}
\usepackage{amsmath}
\usepackage{amsfonts}
\usepackage{graphicx}
\usepackage[numbers]{natbib}
\usepackage{enumitem}
\usepackage{xr}

\usepackage{hyperref}
\usepackage{microtype}
\usepackage{booktabs}
\journalname{Nonlinear Dynamics}
\begin{document}

\title{Sample entropy for graph signals: An approach to nonlinear dynamic analysis of data on networks

}


\author{Mei-San Maggie Lei         \and
        John Stewart Fabila Carrasco \and Javier Escudero 
}


\institute{M.-S.M.L. \and J.E. \at
              Institute for Imaging, Data and Communications, School of Engineering, University of Edinburgh, Scotland, UK.  \\
              \email{M.S.Lei@sms.ed.ac.uk}   \\
            J.S.F.C \at 
		School of Computer Science and Informatics, Cardiff, CF24 4AG, UK. \\ 
		\email{Fabila-carrascoJ@cardiff.ac.uk} \\
            J.E. \\
            \email{Javier.Escudero@ed.ac.uk}
}

\date{Received: date / Accepted: date}

\maketitle

\begin{abstract}
The recent extension of permutation entropy and its derivatives to graph signals has opened up new horizons for the analysis of complex, high-dimensional systems evolving on networks. However, these measures are all fundamentally rooted in Shannon entropy and symbol dynamics. In this paper, we explore, for the first time, whether and how a popular conditional-entropy based measure --Sample Entropy (SampEn)-- can be effectively defined for graph signals and used to characterise the nonlinear dynamics of data on complex networks.

We introduce sample entropy for graph signals (SampEn$_{G}$), a unified framework that generalises classical sample entropy from uni- and bi-dimensional signals, including time series and images, by building on topology-aware embeddings using multi-hop neighbourhoods and computing finite scale of correlation sums in the continuous embedding state space. Experiments on synthetic and real-world datasets verify that SampEn$_G$ recovers known nonlinear dynamical features on paths and grids. In a traffic-flow analysis, SampEn$_{G}$ on a directed topology (encoding causal flow constraint) shows promise to detect phase transitions between free-flow and congestion, offering information that is complementary to existing Shannon-entropy based approaches. We expect SampEn$_{G}$ to open up new ways to analyse graph signals, generalising sample entropy and nonlinear analysis based on conditional entropy to a wide variety of network data.

\end{abstract}

\section{Introduction}
\label{intro}
Technological advances have enabled the large-scale collection of high-dimensional data from complex systems, including traffic systems, financial markets, ecosystems, societal interactions, many of which can be modelled as dynamical systems on networks \cite{RoadTrafficComplexsystem,complex_systems, complex_sys_modelling}. These complex systems often exhibit emergent behaviours, arising from intrinsic interactions and nested interdependencies, which traditional methods may struggle to capture \cite{what_is_complex_system}. 

Such real-world complex signals often exhibit nonlinear behaviour attributed to intrinsic dynamics, while being influenced by noise. It is important to quantify the amount of deterministic and random components--the complexity--of these systems. This motivates the development of nonlinear entropy-based techniques \cite{MVME, DE, RCMPE, biDE}. Among these, Permutation Entropy (PE) and its derivatives, Dispersion Entropy (DE), and Bubble Entropy (BE) have demonstrated their effectiveness in assessing irregularity in numerous applications \cite{entropy_review}. Importantly, these three entropy-based measures have recently been extended for the graph signal domain.

Graph Signal Processing (GSP) has provided a powerful framework to analyse signals (and data in general) with regards to the underlying topology of a system. Each signal value is assigned with a node in a graph, while the edges encode relationships between the agents represented by the nodes. GSP has unlocked a new frontier by extending classical signal processing techniques to data on irregular, non-Euclidean domains \cite{gsp_overview}. In this context, the recent development of those entropy techniques for graph signals -- PE$_G$, DE$_G$, and BE$_G$ -- drastically expand the spectrum of data types to which these classical entropy analysis can be applied to \cite{fabila-carrascoPEG2022,FABILACARRASCO2023113977,bubbleen_graph}. 

While effective, it is noteworthy that the existing methods PE$_G$, DE$_G$, and BE$_G$ are rooted in Shannon entropy, all belonging to the same broad family of non-linear analysis techniques \cite{entropy_review}. In essence, all of them leverage symbols of length $m$ signal samples to compute the source entropy. Other nonlinear analysis frameworks remain to be explored within GSP. 

One major nonlinear metric widely used across a variety of fields is Sample Entropy (SampEn) \cite{ApproximateEntropySample}. SampEn is grounded in similar ideas as the correlation integral: using a distance threshold $\epsilon$ to define a geometric radii (ball) in state space and count how often patterns fall within the neighbourhood. From these correlation integrals, the correlation dimension $D_2$ is obtained by examining how the integrals scale as $\epsilon \rightarrow 0$, providing a measure of the effective fractal dimension of the underlying attractor \cite{grassbergerCharacterizationStrangeAttractors1983}.

SampEn instead utilises a fixed scale $\epsilon$, which is a fraction of the standard deviation of time series. The theoretical approach to information rate would involve the joint probability of arbitrarily long sequences for $m\rightarrow\infty$, which is infeasible in real-world  settings. SampEn computes a low-order, conditional-entropy based measure on the assumption that the marginal distribution in length $m$ patterns remains similar when extended to length $m+1$ \cite{ApproximateEntropySample}. In essence, SampEn computes the correlation sums for patterns of length $m$ and $m+1$ and their ratio to estimate the conditional probability. In this way, SampEn provides a practical -- finite-$m$, finite-$\epsilon$ -- statistic of dynamical irregularity  \cite{sampenfirstdefin}. In contrast to PE, DE and BE, which operate on discretised symbol sequences, SampEn does not impose a symbolic partition on the state space, but works directly in continuous space via a distance threshold.

Thus, SampEn has been frequently used to assess the reccurrence and persistence of similar patterns within a series \cite{sampenfirstdefin}, bringing an auxiliary perspective to nonlinearity assessment from other entropy families. An advantage of SampEn is its relative robustness to short recordings and measurement noise, which is particularly valuable for some real-world signals. Example applications include the analysis of electrocardiography \cite{sampen_ECG}, electroencephalography (EEG) \cite{eeg_sampen_most_used}, and functional magnetic resonance imaging datasets \cite{fmri_short_se}. Notably, SampEn is the most commonly used entropy metric for neurological disorder detection in EEG signals for the past decade \cite{eeg_sampen_most_used}. Despite its origin in medicine, SampEn has been applied to other fields including finance, electronics, ecology, and engineering \cite{ApproximateEntropySample}. 

In this paper, we define SampEn for graph signals, thereby providing a graph-based complexity measure for the characterisation of dynamic processes evolving on networks. The proposed measure, SampEn$_{G}$ reduces to classical unidimensional (1D) SampEn \cite{sampenfirstdefin} on path graphs, generalises two-dimensional (2D) SampEn \cite{sampen_2d} on grid graphs, allowing us to quantify dynamics of signals defined on arbitrary network topologies. Our contributions are:
\begin{itemize}
    \item The first generalisation of SampEn to graph signals (SampEn$_{G}$) that works regardless of whether the graphs are directed or undirected, binary or weighted, providing the first conditional-entropy based nonlinear measure for graph signals. 
    \item Experiments confirming that SampEn$_{G}$ replicates the behaviour of classical SampEn for time series and SampEn$_{2D}$ for images when applied to appropriate regular graphs, and illustrating its application to evaluation of dynamics on graph signals.
    \item A study of the dependency of SampEn$_{G}$ on its parameters and its behaviour in graph construction and the presence of noise.
\end{itemize}

\section{Background}
\label{sec:guidelines}

\subsection{Classical SampEn}
\label{SampEn_classic}

SampEn quantifies the likelihood of repeatability within a time series by examining a correlation integral-like count of matched patterns of length $m$ that remain similar when extended to length $m+1$ \cite{sampen_typical_param}. Two patterns are considered a ``match'' if they have a Chebyshev distance less than a tolerance threshold: $\epsilon = r\times SD$, where $SD$ is the standard deviation of the time series.  Typical values for the parameters of SampEn are embedding dimension $m=\{1,2\}$ and tolerance $r \approx 0.2$ \cite{sampen_typical_param}. 

Given a 1D signal \(\mathbf{x} = \{x_i\}_{i=1}^{i=N}\), and parameters $m$ and $r$, classical SampEn is computed as follows:

\begin{enumerate}[wide, labelwidth=!, labelindent=0pt, itemsep=1pt]

    \item For \(i = 1,\dots, N-m\), consecutive, overlapping patterns of length $m$ and $m+1$ can be constructed with a sliding window on the signal $\mathbf{x}$ as:%
        \begin{equation*}
            \mathbf{x}_m(i) = [x_i, x_{i+L}, \dots,x_{i+L(m-1)}],
        \end{equation*}
        \begin{equation*}
            \mathbf{x}_{(m+1)}(i) = [x_i, x_{i+L}, \dots,x_{i+Lm}].
        \end{equation*}
        $L$ is a time step that separates the samples and it is often taken as $L = 1$. Hence, it will be omitted from the rest of the formulation.
    \item \label{similarity_m} Count matching patterns for embedding dimension $m$, $B_i$ as the evaluation of a comparison between the pairwise Chebyshev distance and the pre-defined tolerance threshold ($r\times SD$). For $1\leq j \leq N-m$ and $j\neq i$:
    \begin{align}
    B_i(r) &= 
    \tfrac{1}{N - m-1} \sum_{\substack{j = 1, \\ j \neq i}}^{N - m} 
    \| d[\mathbf{x}_m(i), \text{ } \mathbf{x}_m(j)] \leq (r\times SD) \|,
    \label{eq:B_match_each}
    \end{align}
    where $\|\cdot\|$ denotes cardinality of a set and the Chebyshev distance for each pair of patterns is by:
    \begin{equation}
    d[\mathbf{x}(i), \mathbf{x}(j)] = \max_{k=0,\ldots,m-1}|x(i+k) - x(j+k)|.
    \label{eq:chebyshev}
    \end{equation}
    We compute the correlation sums of $B_i$ for all $N-m$ patterns excluding self-match to give the global $B^m$ count for embedding dimension $m$:    
    \begin{equation}
    B^m(r) = \frac{1}{N-m} \sum_{i=1}^{N-m} B_i(r). 
    \label{eq:B_match_total}
    \end{equation}

    \item Step \ref{similarity_m}) is repeated for embedding dimension $m+1$ to count matches $A_i$ for each \(\mathbf{x}_{m+1}(i)\) and the averaged total count $A^m$, using the same tolerance (\(r\times SD\)):
    \begin{equation}
    A^m(r) = \frac{1}{N-m} \sum_{i=1}^{N-m} A_i(r) \label{eq:A_match_total}.
    \end{equation}
    \item Finally, we compute SampEn as the ratio of $A^m$ and $B^m$:
        \begin{equation}
            \text{SampEn}(m, r, N) = -\ln\left(\frac{A^m(r)}{B^m(r)}\right).
            \label{eq:sampenapproxfinal}
        \end{equation}
    
\end{enumerate}

Importantly, SampEn has been extended to images (2D data) \cite{sampen_2d}. In contrast to the 1D scenario, the 2D version considers overlapping square windows of size \(m \times m\) and \( \left(m+1\right) \times \left(m+1 \right) \)  from the image, creating $(N - m)\times(N - m)$ patterns. The Chebyshev distance is adapted to measure the pairwise difference of the corresponding pixels between two patterns centred at pixel $(i_1,i_2)$ and $(j_1,j_2)$. The subsequent steps to compute SampEn$_{2D}$ follow the univariate SampEn algorithm as Step 2-4) in Sec.~\ref{SampEn_classic}.

\subsection{Graphs and graph signals}
\label{SampEn_graph_ext}

In this study, we generalise SampEn to graph signals. As preliminaries, we first define a graph and a graph signal. A graph is formulated by \(\mathcal{G} = (\mathcal{N}, \mathcal{E}, \mathbf{A})\), where $\mathcal{N}$ and $\mathcal{E}$ denote the set of $N$ nodes and edges correspondingly, and $\mathbf{A} \in \mathbb{R}^{N\times N}$ is the adjacency matrix: a square matrix encoding the edge connectivity and weights between the nodes \cite{big_data_graph}. A directed, unweighted adjacency matrix $\mathbf{A}$ can be populated as a binary matrix:
\[
	\textbf{A}_{i j}= 
	\begin{cases}
		1 & \text{if there exists a edge from node } i \text{ to } j,  \\
		0 & \text{otherwise,}
	\end{cases}
	\label{eq:Adjacency}
\]
 for $i,j=1,\dots,N$. For weighted graphs, $\textbf{W}$ is the weighted adjacency matrix, each entry $\textbf{W}_{i j}$ is the weight \( w_{ij} > 0 \) if an edge exists, whereas $\textbf{A}_{\mathrm{undir}} = \textbf{A}_{\mathrm{dir}} \cup \textbf{A}_{\mathrm{dir}}^\top,
$ is a symmetric matrix in the case of an undirected graph.

Graph signals represent various types of data depending on the application by assigning to each node on the graph a signal value \( \mathbf{x} = [x_1, x_2, \dots, x_N]^\top \in \mathbb{R}^{N \times 1}\). For instance, in a sensor network, each node corresponds to a sensor, and the signal value can represent a measurement such as temperature or pressure \cite{big_data_graph}.

\section{SampEn\texorpdfstring{$_G$}{\_G}}\label{Sec:sampengraph}

In the classical SampEn for univariate time series (hereafter refer to as 1D SampEn too), the patterns are built by concatenating successive temporal observations separated by the time step $L$ (often $L=1$). In our graph-based generalisation, time is replaced by graph hop distance --  a hop is a single traverse along an edge from one node to another. 

For each node, we construct a vector which components represents the signal samples in steps of increasing $L$ hops (0, $L$, $2L$, $\dots$). We then aggregate over multi-hop neighbourhoods to characterise the radial profiles around each node via powers of the graph shift operator (adjacency matrix), in analogy to increasing temporal steps in SampEn.

To obtain these components, we mathematically identify the local $L$-hop neighbours for each node. Consider the multiplication of $\mathbf{A}$ with itself: 
\[
    (\mathbf{A}^{2})_{ij} = \sum_{k} \mathbf{A}_{ik} \times \mathbf{A}_{kj}
\]
Each entry $(\mathbf{A}^{L=2})_{ij}$ is the sum of walks of length \textit{two} from node $i$ to $j$ over any intermediate node $k$. The $L$-hop degree of node $i$ can be defined as the corresponding row sum:
\begin{align}
    \deg^{L}(i) &:= \sum_{j=1}^N (\mathbf{A}^L)_{ij}, \qquad L\ge 1.
    \label{eq:degL}
\end{align}

We identify the $L$-neighbourhood, nodes reached through L-step walks, resulting in a walk-weighted L-step neighbourhood, of each node. For $L\in\mathbb{N}$, non-zero entries in \((\mathbf{A}^L)_{ij}\) indicate the total number of $L$-length walks reachable from node $i$ to $j$.

Building on our previous work on PE$_G$ \cite{fabila-carrascoPEG2022}, we here extend the framework to introduce SampEn$_{G}$. For each node $i=1,\dots,N$, we define its $m$-dimensional graph-aware patterns $\mathbf{\overline{x}}^{m}(i)$ as:
\begin{equation}
    \bar{\mathbf{x}}^{m,L}(i)=[\bar{x}_i^0,\bar{x}_i^L,\ldots,\bar{x}_i^{(m-1)L}],  \quad i \in \mathcal{N}^\star_{m},
\end{equation}
\begin{align}
    \mathcal{N}^\star_m := \big\{ i\in\mathcal{N} \;:\; \deg^{L}(i)>0 \;\; \forall\, L=1,\dots,m \big\}.
    \label{eq:Vstar-final}
\end{align}

To avoid undefined patterns with sparse graphs, we restrict to patterns with non-zero reachability up to $m$-hops. Here, $\overline{x}_i^0$ is the signal value on node $i$, and $\overline{x}_i^L$ is a walk-weighted mean of the $L$-neighbourhood of node $i$, in hop-radius propagation:
\begin{equation}
    \quad \overline{x}_i^{L} = \frac{1}{\text{deg}^{L}(i)} \sum_{j \in \mathcal{N}_{L}(i)} (\mathbf{A}^{L})_{ij}x_j.
    \label{eq:segraphavg}
\end{equation}

\(\mathcal{N}_L(i)\) denotes the set of nodes reachable by at least one length-$ L $ walk from node \(i\). For weighted graphs, we replace \(\mathbf{A}\) with the weighted adjacency \(\mathbf{W}\). The matrix power $\mathbf{W}^L$ is the total weighted walks of $L$ hops from $i$ to $j$: \(\text{deg}^{L}(i) = \sum_{j=1}^N (\mathbf{W}^L)_{ij}, 
    \quad \overline{x}_i^{L} = \frac{1}{\text{deg}^{L}(i)} \sum_{j \in \mathcal{N}_{L}(i)} (\mathbf{W}^{L})_{ij}x_j.
\)

The resulting patterns are of size $1\times m$ for each node and form the basis for the subsequent steps in constructing the $m+1$ patterns in an analogous way. The successive steps of similarity matching, then follows:
\begin{equation}
    B_i(r) = \frac{1}{|\mathcal{N}^\star_{m}|-1}\!
    \sum_{\substack{j\in\mathcal{N}^\star_{m} \\ j\neq i}}
    \| d[\mathbf{\overline{x}}^{m}(i), \text{ } \mathbf{\overline{x}}^{m}(j)] \leq \epsilon \|,
    \label{eq:Bi-graph}
\end{equation}
with \(\epsilon=r\times SD\) as per the classical definition, the overall correlation sums are then computed over nodes to obtain $B^m(r)$: 
\begin{equation}
    B^m(r) = \frac{1}{|\mathcal{N}^\star_{m}|} \sum_{i\in\mathcal{N}^\star_{m}} B_i(r). 
    \label{eq:B_match_total_graph}
\end{equation}

Repeating for $(m+1)$-patterns yields $A^m(r)$. The computation of SampEn$_{G}$ follows as Eq.~\eqref{eq:sampenapproxfinal}.

\subsection{Choice of $m$, $r$, and $L$}\label{Sec:Choiceofparam}
The parameters have complementary roles. The embedding dimension $m$ controls how many graph-propagation scales are included in each pattern; the tolerance $r$ fixes the amplitude scale at which two patterns are regarded as recurrent; and the hop-lag $L$ controls the spatial or relational separation between successive components of the graph pattern. In all experiments below, unless otherwise stated, we use $L=1$ so that the embedding uses the immediate graph dynamics and remains directly comparable with classical SampEn on a directed path.

Increasing $m$ or $L$ increases the largest explored hop depth $mL$. This can be useful when the dynamics propagate beyond immediate neighbours, but it also increases neighbourhood overlap. If the graph is dense, small-world, or contains many long-range shortcuts, the $qL$-step neighbourhoods may rapidly cover a large fraction of the graph. Then the walk-weighted averages $\bar{x}_i^{qL}$ become similar across nodes and $A^m(r)/B^m(r)$ approaches one, causing $\mathrm{SampEn}_G$ to decrease or plateau. Therefore, for dense graphs or graphs with large hubs, we recommend $L=1$ and small embeddings, typically $m=1$ or $m=2$. For sparse lattices, paths, sensor chains, or road networks, $L=1$ remains the default, while $L=2$ can be used as a sensitivity check when interactions are expected to occur over a wider neighbourhood.

A practical diagnostic is to monitor the effective valid sample size $M=|\mathcal{N}_{m,L}^{\star}|$ and the mean relative reachability
\begin{equation}
    \rho_h=\frac{1}{N}\sum_{i=1}^{N}\frac{|\mathcal{N}_h(i)|}{N},\qquad h=mL.
\end{equation}
Very small $M$ indicates that the graph is too disconnected for the chosen $mL$, whereas $\rho_h$ close to one indicates that most node patterns use nearly global information. In either case, $m$ or $L$ should be reduced. In line with classical SampEn practice, we use $m\in\{1,2\}$ as the main range and report sensitivity over $r\in[0.1,0.25]$; $r\approx0.2$ is a useful default, but the most reliable operating region is where the qualitative separation between regimes is stable over a range of $r$ values.

\subsection{Computational complexity}\label{par:complexityanalysis}
Let $N$ be the total number of nodes in the graph and let $E$ be the number of nonzero edges. For each valid node, SampEn$_G$ builds one graph-embedding vector. We denote by
\begin{equation}
    N_{\mathrm{valid}} = |\mathcal{N}_{m,L}^{\star}|
\end{equation}
the number of valid embedding vectors used in the SampEn$_G$ calculation. Since each valid node contributes one embedding vector, $N_{\mathrm{valid}}\leq N$.

The computation has two main steps. First, the graph-averaged values
\begin{equation}
    \bar{x}_i^0,\bar{x}_i^L,\bar{x}_i^{2L},\ldots,\bar{x}_i^{mL}
\end{equation}
are computed for all valid nodes. These values are needed because SampEn$_G$ compares both $m$-dimensional and $(m+1)$-dimensional graph patterns. The largest required hop depth is therefore $mL$. If these graph averages are computed by repeated sparse graph propagation, the cost of this step is
\begin{equation}
    O(mLE).
\end{equation}

Second, SampEn$_G$ compares all pairs of valid graph-embedding vectors. There are approximately $N_{\mathrm{valid}}^2$ such pairs, and each comparison costs $O(m)$ because the vectors have length $m$ or $m+1$. Therefore, the exact pairwise matching step costs
\begin{equation}
    O(mN_{\mathrm{valid}}^2).
\end{equation}

Combining the two steps, the total computational complexity is
\begin{equation}
    T_{\mathrm{SampEn}_G}
    =
    O(mLE + mN_{\mathrm{valid}}^2).
\end{equation}

For fixed $m$ and $L$, and for sparse graphs where $E=O(N)$, the main bottleneck is the pairwise matching term $O(mN_{\mathrm{valid}}^2)$. Thus, the exact SampEn$_G$ algorithm scales quadratically with the number of valid nodes. The memory cost is $O(mN_{\mathrm{valid}})$ if pairwise distances are computed block by block, without storing the full distance matrix.

\begin{figure*}[h]
    \centering
    \includegraphics[width=.9\textwidth]{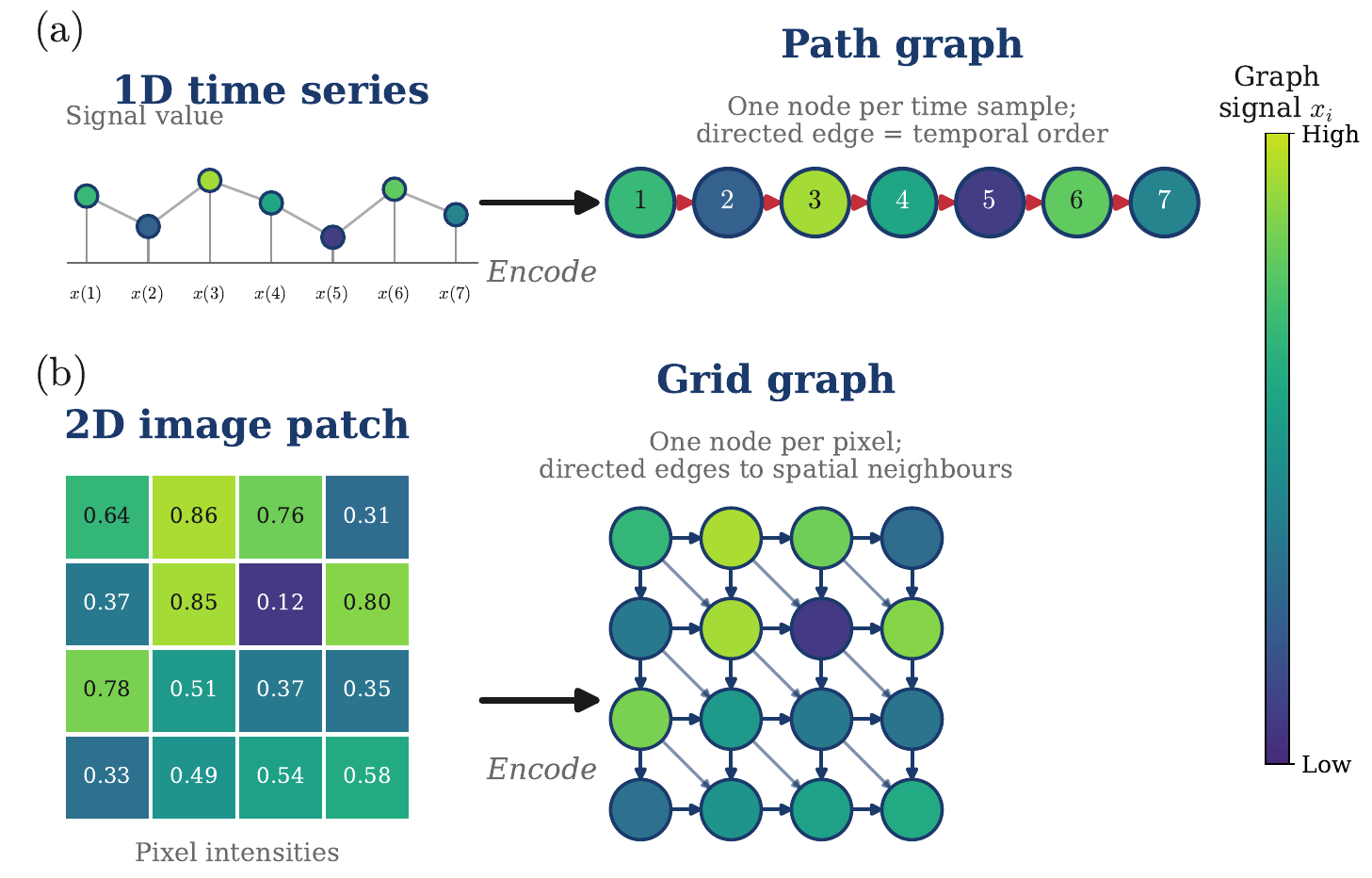}
    \caption{Encoding signals as graphs: (a) a 1D time series on a directed path (node = time sample, edge = temporal order); (b) a 2D image patch on a grid (node = pixel, edges to spatial neighbours). Colour shows the signal value $x_i$. These recover 1D SampEn and SampEn$_{2D}$ as special cases.}
    \label{fig:demon_1d2d}
\end{figure*}

\subsection{SampEn$_{G}$ reduces to 1D SampEn on a path graph} 
It has been shown that a time series can be modelled as a graph signal on a directed path, and the graph-based embedding used in PE$_G$ reduces to the 1D embedding in classical PE (see Proposition~3 in \cite{fabila-carrascoPEG2022}). 

We adopt the same multi-hop pattern construction for SampEn$_{G}$ but apply the SampEn functional instead. On directed path graphs, SampEn$_{G}$ reduces to the classical SampEn defined in Sec.~\ref{SampEn_classic}, a property that we numerically validate in Sec.~\ref{Sec:logistic}. See Fig.~\ref{fig:demon_1d2d}.

\subsection{The behaviour of SampEn$_{G}$ aligns with that of 2D SampEn on regular grids}
When the underlying graph is an 8-neighbour regular grid with pixels as nodes, the radial pattern of multi-hop averages $ \overline{x}_i^L $ over $ L $-hop neighbourhoods for each node summarises the local image structure at increasing spatial scales. Although these embeddings are not pixel-wise identical to the square patches of SampEn$_{2D}$ introduced in \cite{sampen_2d}, we empirically evaluate the behaviour of SampEn$_G$ acting as a natural graph-based analogue of SampEn$_{2D}$ on image-like domains (Sec.~\ref{Sec:Exp_images2D}).












\section{Experiments and results}
\label{Exp_setup}


We evaluate SampEn$_{G}$ on synthetic and real-world datasets to validate:
\begin{enumerate}
    \item Generalisation ability to classical SampEn on path graphs (1D), and alignment with SampEn$_{2D}$ on grid graphs.
    \item Sensitivity to transition in structure (signal and topology) under low and high-noise regime, and parameter choices.
    \item Practicality as a dynamics index on real-world graph signals.
\end{enumerate} 

Synthetic datasets include 1D logistic map, Brodatz textures, MIX$_{2D}$ process, Erd\H{o}s--R\'enyi (ER) random graphs, and Watts-Strogatz (WS) small-world networks.

Real-world applications include a weather station dataset \cite{giraultStationaryGraphSignals2015}, the Intel Berkeley sensor dataset \cite{intelLabData2004}, and a freeway traffic dataset \cite{coursey2024ftaed}.

We set embedding dimension \(m\) and tolerance parameter \(r\) to typical values used in classical SampEn applications: \(m\in[1,2,3]\) and \(r\) in the range of \(0.1\) to \(0.25\) \cite{pincusAssessingSerialIrregularity2001} and report mean $\pm$ standard deviation across 20 repetitions for all synthetic experiments. The code used in this study is publicly available at: \url{https://github.com/mslmaggie/SampEnG}.

\subsection{Logistic map (1D)}
\label{Sec:logistic}

We first validate SampEn$_{G}$ on the logistic map, a classical nonlinear dynamical model where the system oscillates between order and chaos as a function of its bifurcation parameter $\rho$. The map is defined by \cite{logistic_map_sampen}: \begin{equation}
    x_{t+1} = \rho x_t (1 - x_t).
\end{equation}

The time series is represented as a graph signal on a binary path. Each time sample is associated to a node with directed \(i\rightarrow i+1\), or undirected edges between consecutive nodes, following the protocol in \cite{fabila-carrascoPEG2022}. We compute SampEn$_{G}$ for both paths and evaluate them against classical 1D SampEn (implementation from \cite{MartinezCagigal2018SampleEntropy}) in relation to \(\rho\). The experiment is repeated over 20 random initial conditions drawn uniformly from the open interval \((0,1)\), and we report the mean and standard deviation across runs.

\begin{figure*}[h!]
    \centering
    \includegraphics[width=\textwidth]{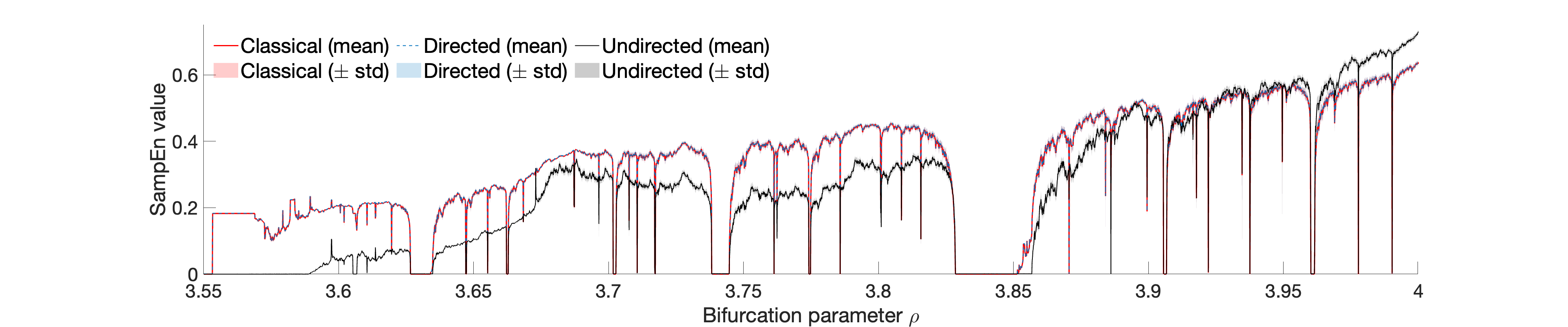}
    \caption{Mean and standard deviation reported for: Directed (blue), undirected (black) SampEn\(_{G}\), classical SampEn (red) computed for \(m=2\), \(r=0.2\) against $\rho$ for logistic map. Repeated over 20 random initial conditions drawn uniformly from \((0,1)\).}
    \label{fig:logisticmap}
\end{figure*}

Fig.~\ref{fig:logisticmap} shows that SampEn$_{G}$ on the directed path replicates the classical SampEn results, confirming the proposed formulation reduces to the standard 1D case on path graphs. This is because the directed graph imposes the causality in the order of samples within the patterns of length $m$ and $m+1$ implicitly assumed by classical SampEn. The undirected-path SampEn$_{G}$ closely tracks the overall trend of the other two curves, correctly identifying the bifurcation points and islands of periodicity with slight quantitative differences due to the non-causal, symmetric neighbourhood structure of the undirected path. These results validate the generalisation property of SampEn$_{G}$ on 1D signals. 

\subsection{Image datasets (2D)}
\label{Sec:Exp_images2D}
We next assess the behaviour of SampEn$_G$ on images in comparison to results from SampEn$_{2D}$ on Brodatz texture analysis, and its sensitivity to structured content under noise-dominated regime using the MIX$_{2D}(p)$ process, in line with \cite{sampen_2d}. 

In both cases, each image is represented as a graph signal on a directed, binary 8-neighbour grid where each node is a pixel, edges encode the 8-neighbour connectivity, and signal values are the pixel intensities  associated to their corresponding nodes, following \cite{fabila-carrascoPEG2022}.

\subsubsection{Brodatz Texture dataset}
\begin{figure}[t!]
    \centering
    \includegraphics[width=0.65\linewidth]{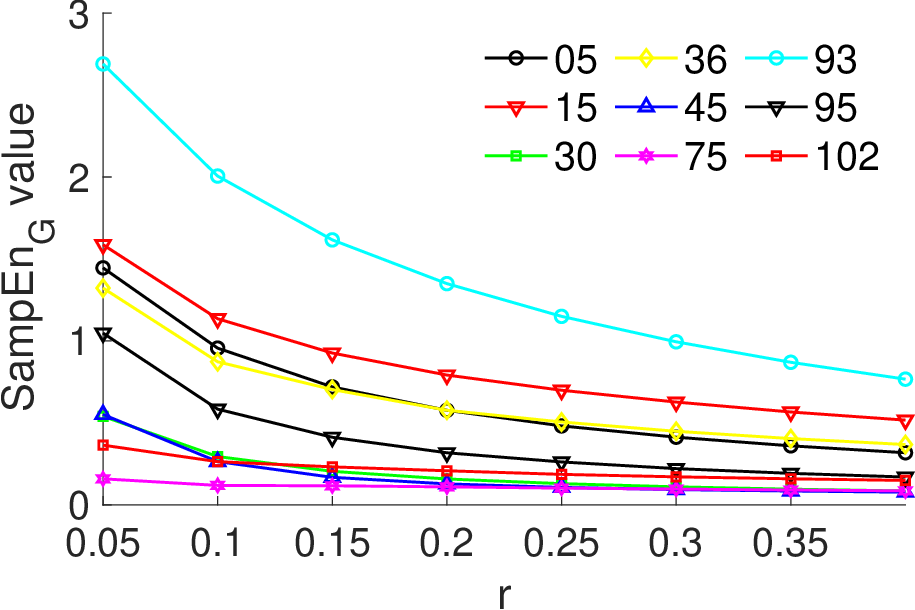}
    \caption{Mean SampEn$_{G}$ against $r$ for 9 textures groups ($m=2$). Values are averaged over 25 non-overlapping \(128\times128\) patches for each texture in the Brodatz dataset.}
    \label{fig:brodatz}
\end{figure}

The Brodatz dataset consists of \(640\times640\) grayscale images for irregularity and texture analysis \cite{texture_eval}. We partitioned selected images of nine pattern groups into 25 non-overlapping patches, each $128\times128$, for analysis on par with \cite{sampen_2d}. For each patch, we compute SampEn$_{G}$ as a function of $r$. The averaged results across patches for each group are displayed in Fig.~\ref{fig:brodatz}.

Across texture groups, the relative ordering of SampEn$_{G}$ results as $r$ varies mirrors those reported for SampEn$_{2D}$ in \cite{sampen_2d}, where more regular or periodic textures exhibit lower entropy values; and more irregular textures exhibit higher entropy. 

We observe that absolute values from SampEn$_{G}$ is generally lower than those from SampEn$_{2D}$, which difference is likely attributed to a larger set of patterns in formulating each pixel as a node on the graph, which increases the empirical matching probabilities for SampEn$_{G}$. Despite the offset, the overall rankings and primary findings across texture groups and the functional dependence on $r$ are consistent with SampEn$_{2D}$. These findings support the transferability of SampEn$_{G}$ on 2D image analysis.

\subsubsection{MIX$_{2D}(p)$ process}
\begin{figure}[t!]
    \centering
    \includegraphics[width=0.7\linewidth]{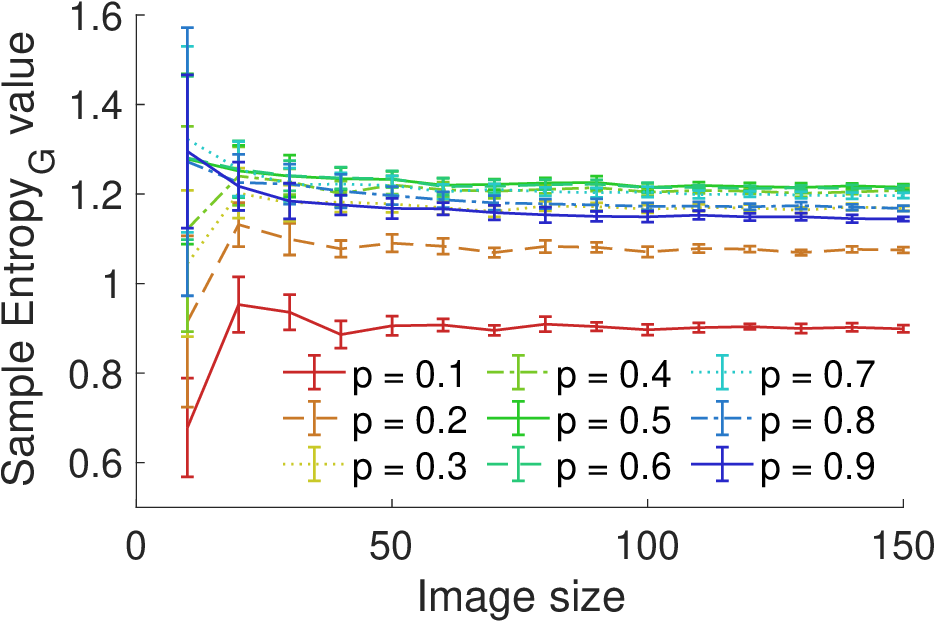}
    \caption{Averaged SampEn$_{G}$ (over 20 repetitions) as a function of image size for various values of noise parameter $p$ for MIX\(_{2D}\) process (\(m=2,r=0.2\)).}
    \label{fig:MIX_res}
\end{figure}

To study the impact of additive noise and image (signal) size on a structured signal and topology, we used the MIX$_{2D}$ process to generate synthetic images under a controlled setting. The pixel intensities are defined by a 2D periodic sine pattern corrupted by additive uniform distributed noise, with overall noise level governed by probability $p$ as in \cite{sampen_2d}. We vary the image size from \(10\times 10\) to \(150\times 150\), and $p$ from $0.1$ to $0.9$ in steps of 0.1. Corresponding results, ran over 20 independent realisations, are reported in Fig.~\ref{fig:MIX_res}.

With moderate noise levels (\(p\leq0.5\)), SampEn$_{G}$ increases with $p$ as regularity becomes gradually overruled by randomness, reflecting progressive loss of signal structure. Stability of our measure also improves significantly with the image size for \(N>10^2\) (\(10\times 10\)) -- variance across realisations decreases, per the literature \cite{sampenfirstdefin}.

At high noise (\(p>0.5\)), the curves become less separable and saturate, for close values of $p$. This behaviour is attributed to two factors. First, when noise dominates the signal, local patterns are primarily governed by the noise distribution rather than the underlying structure. Second, the averaging over multi-hop neighbourhoods operation in pattern formation acts as a low-pass filter, which suppresses high-frequency fluctuations. While this improves robustness to moderate noise on structured signals, it also attenuates informative variations when the signal is highly random at large $p$. 

The local signal patterns then become statistically similar across the image, causing the corresponding neighbourhood-averages to concentrate around a common value, hence matching probability of the patterns falling within the geometric distance $\epsilon$ stabilises and appears more predictable. As a result, the estimated entropy saturates and even decreases slightly in further increase of randomness $p$ in this noise-dominated regime.

This behaviour diverges from those of PE$_{G}$ on MIX$_{2D}$ reported in \cite{fabila-carrascoPEG2022}, where entropy continues to increase with $p$. This reflects the different foundation of SampEn in comparison with PE. In particular, PE$_{G}$ counts ordinal pattern and does not rely on a distance threshold. Every embedding contributes to the pattern statistics, and greater noise produces a more uniform distribution of permutations. However, in SampEn$_{G}$, matches are not guaranteed as they depend on the Chebyshev distance and the tolerance threshold. Under high-noise influence with little to no detectable underlying structure, the entropy measure no longer increases with $p$. Additional embedding dimensions $m$ to $m+1$ only induce small differences in the number of matches, consequently, the conditional unpredictability of patterns decreases.

\subsection{Experiments on synthetic graphs and network dynamics}

Following experiments on classical signals (i.e., 1D time series and 2D images), we analyse SampEn$_{G}$ on synthetic graph models to evaluate its robustness with respect to topology transitions. We consider ER random graphs, which allow us to vary the graph density via an edge probability $p$, and WS small-world networks, which transition from regular lattices to random graphs via a rewiring probability $\beta$. These experiments explore the behaviour of SampEn$_{G}$ when both the signal and topology are random (Sec.~\ref{Sec:ER}) or some arbitrary signal (Sec.~\ref{Sec:WS}), its ability to quantify dynamics as the underlying network become denser or more randomly connected, and its dependency on the choice of pattern length $m$ (with $r=0.2$ fixed). To move beyond random/static graph signals, we next consider the Kuramoto model of coupled oscillators--a controlled nonlinear dynamical system on three different synthetic topologies, used to evaluate whether SampEn$_{G}$ tracks organisation changes such as local-to-global synchronisation in Sec.~\ref{Sec:Kuramoto}.

\subsubsection{ER random graphs}
\label{Sec:ER}
We consider directed ER random graphs with \(N\in\{30, 100, 300, 900, 2700\}\). Each node value is a random sample from a uniform distribution in the range $[0.01,0.10]$. The edge connectivity parameter $p$ determines the probability in which an edge exists between any two nodes. 

\begin{figure}[h!]
    \centering

    \begin{subfigure}[b]{0.33\textwidth}
        \centering
        \includegraphics[width=\textwidth]{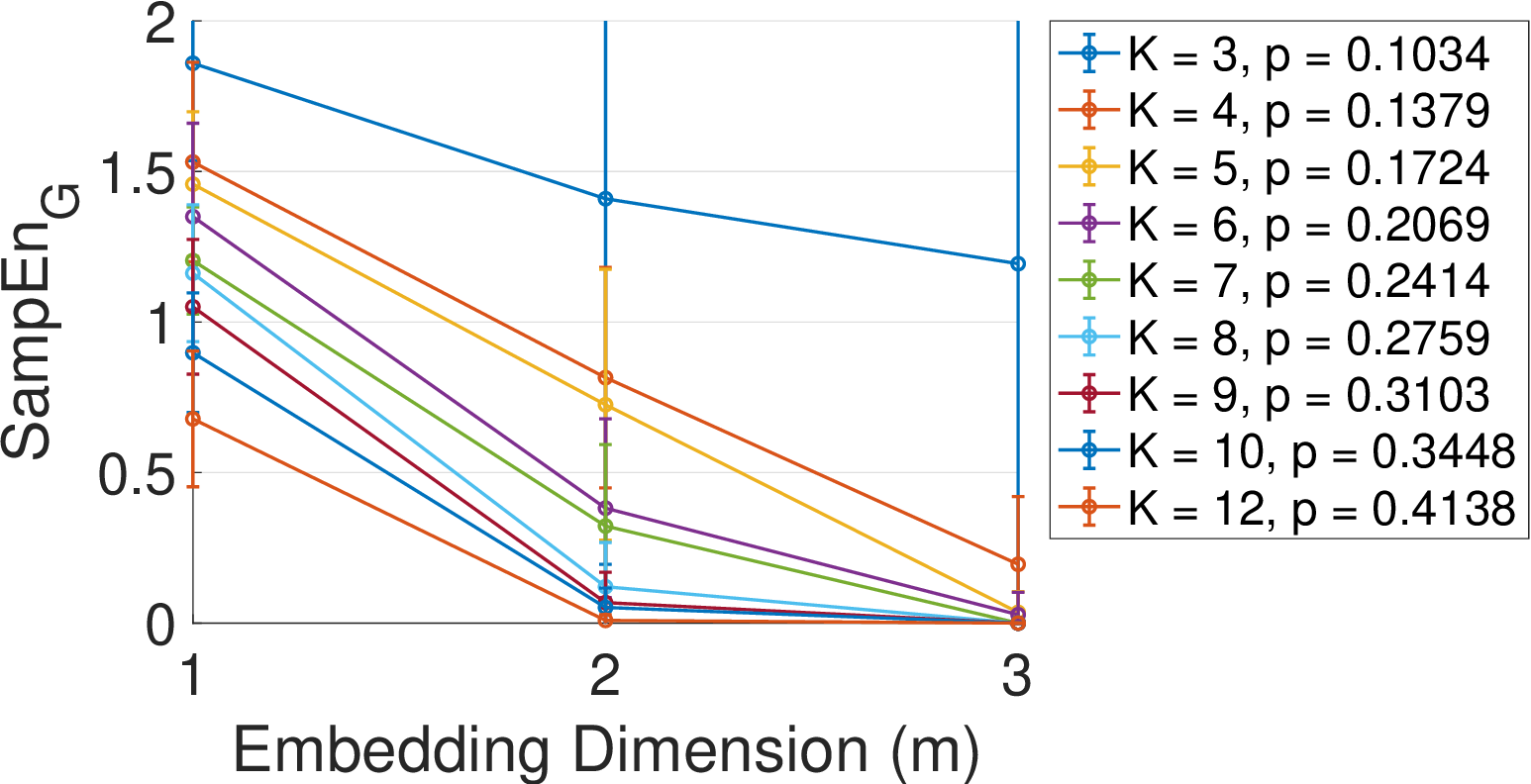}
        \caption{\(N=30\)}
        \label{fig:ERfig1}
    \end{subfigure}
    \begin{subfigure}[b]{0.33\textwidth}
        \centering
        \includegraphics[width=\textwidth]{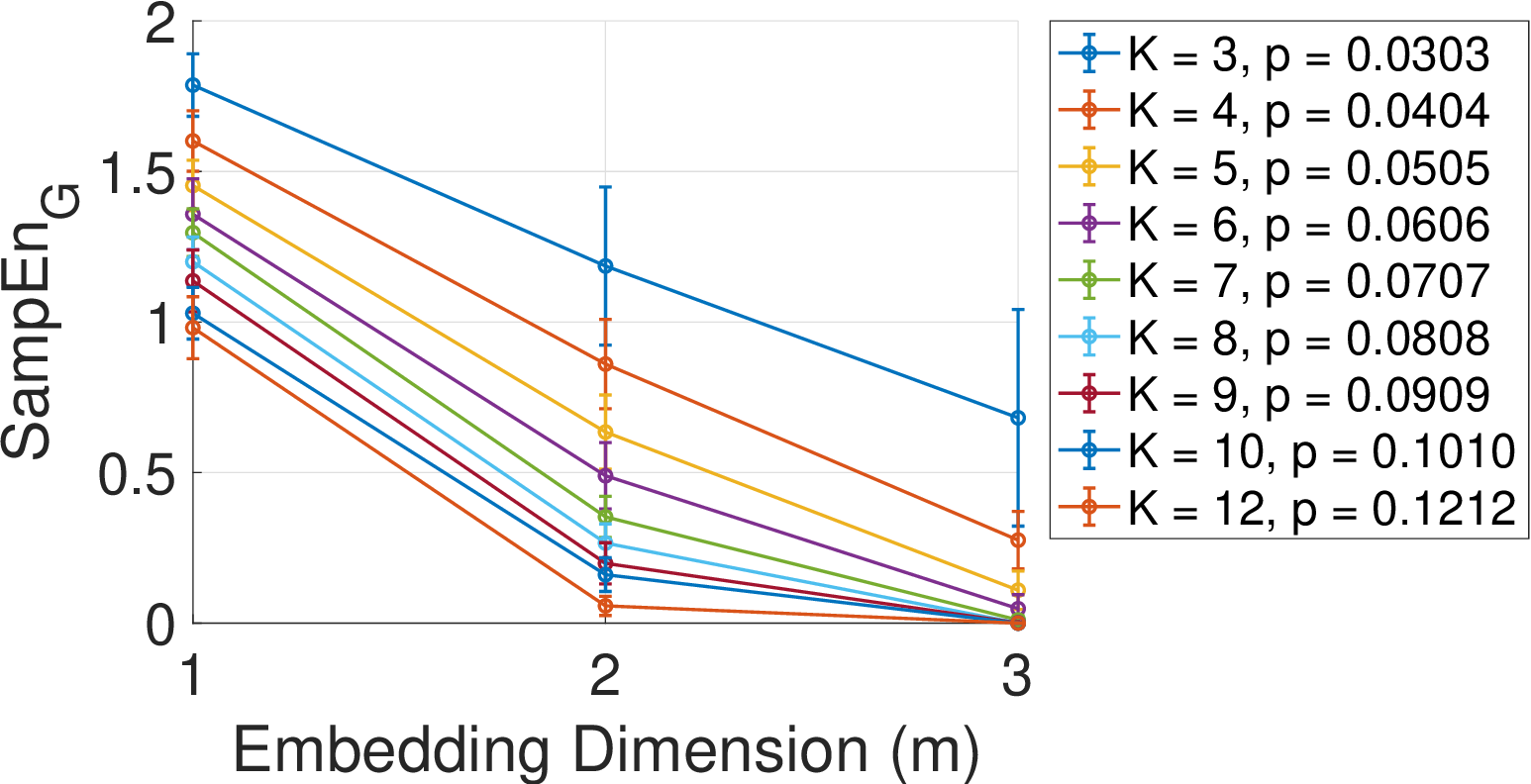}
        \caption{\(N=100\)}
        \label{fig:ERfig2}
    \end{subfigure}
    \hfill
    \begin{subfigure}[b]{0.33\textwidth}
        \centering
        \includegraphics[width=\textwidth]{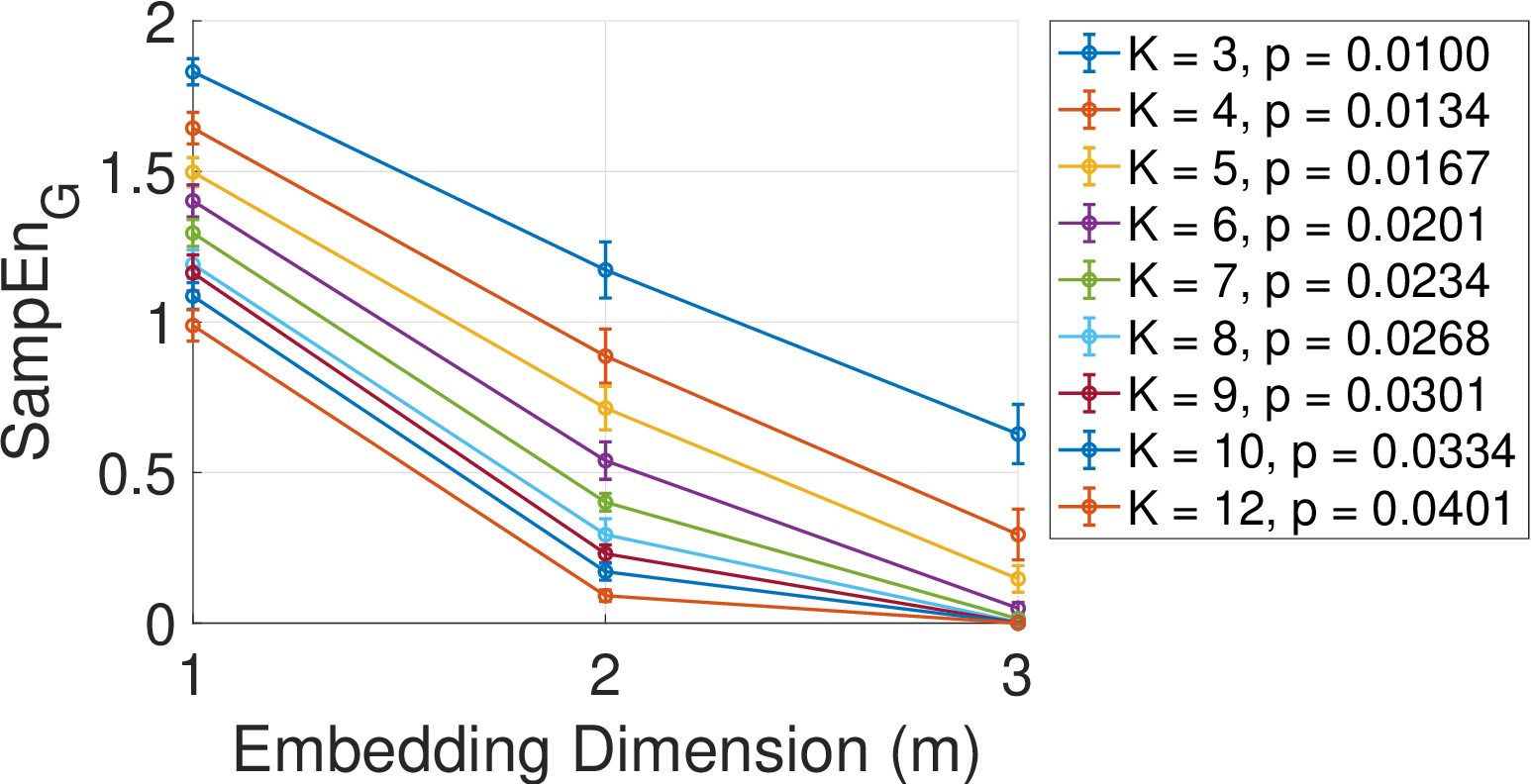}
        \caption{\(N=300\)}
        \label{fig:ERfig3}
    \end{subfigure}
    \hfill
    \begin{subfigure}[b]{0.33\textwidth}
        \centering
        \includegraphics[width=\textwidth]{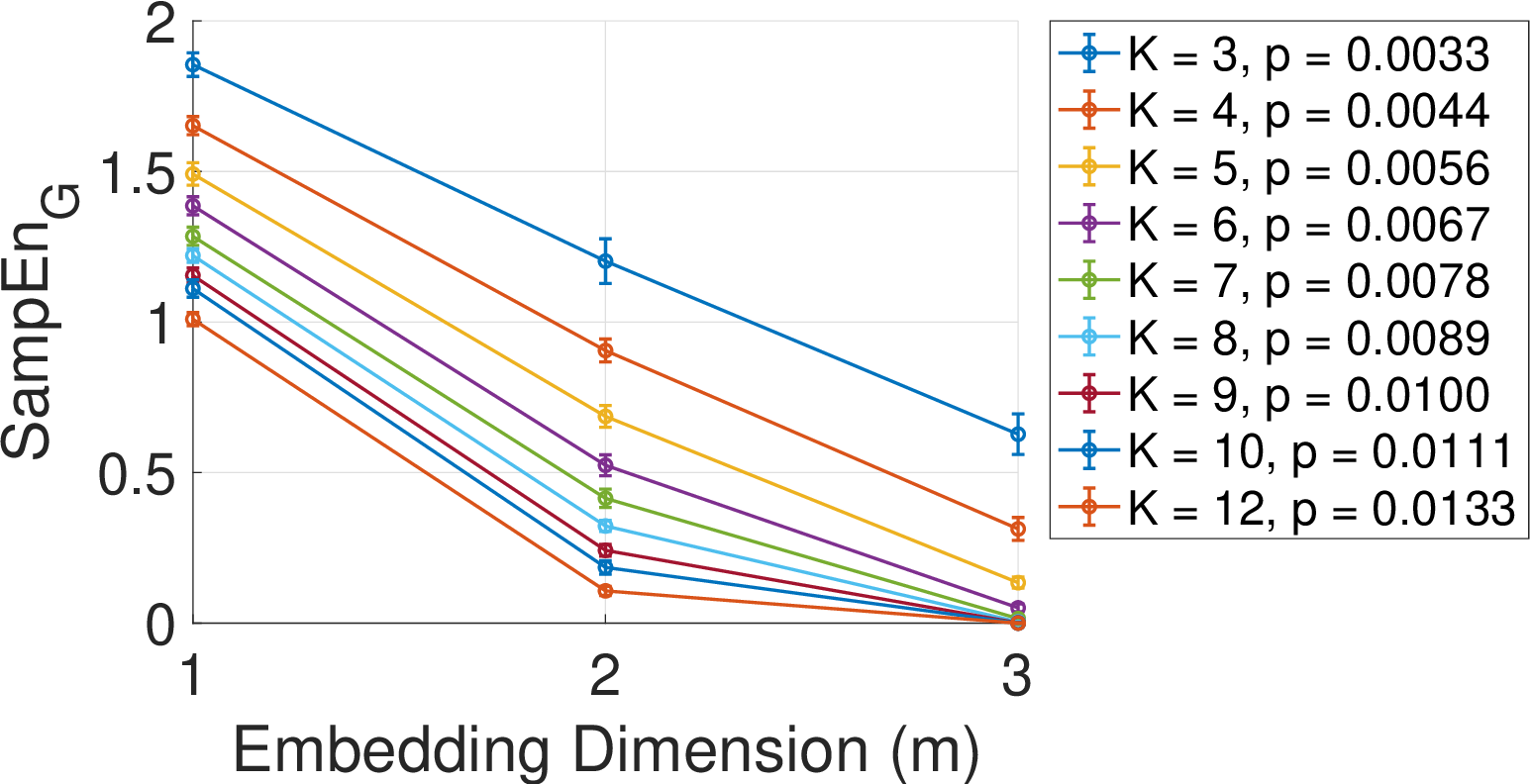}
        \caption{\(N=900\)}
        \label{fig:ERfig4}
    \end{subfigure}
    \hfill
    \begin{subfigure}[b]{0.33\textwidth}
        \centering
        \includegraphics[width=\textwidth]{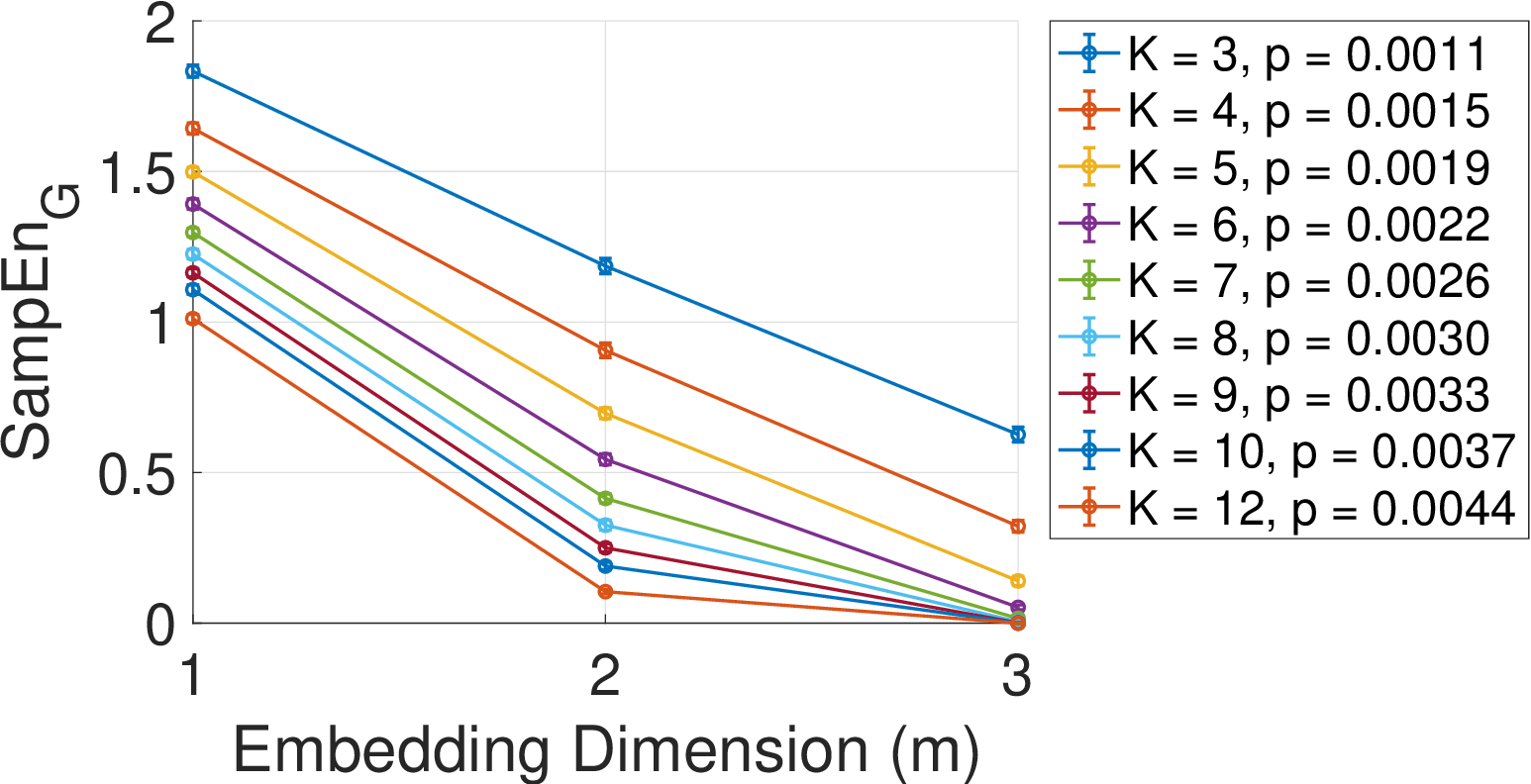}
        \caption{\(N=2700\)}
        \label{fig:ERfig5}
    \end{subfigure}

\caption{$ \mathrm{SampEn}_G $ on directed Erd\H{o}s--R\'enyi graphs with $ N\in\{30,100,300,900,2700\}$ nodes: mean $\pm$ standard deviation over 20 realisations versus connectivity parameter $p$ (targetting mean out-degree $ K$) for embedding dimensions $ m\in\{1,2,3\} $ with tolerance $ r=0.2 $.}

    \label{fig:sampen_ER_res}
\end{figure}

We compute the corresponding $p$ values to achieve a target mean out-degree \(K\in\{3,4,5,6,7,8,9,10,12\}\) using:
\begin{equation}
    p = \frac{K}{N-1},
    \label{eq:ERp}
\end{equation}
for each graph size $N$. For each combination of (\(N,p\)), we evaluate SampEn$_{G}$ as a function of pattern length $m \in [1,2,3]$ over 20 graph realisations. We simultaneously measure the mean computation time for each (\(N,K,m\)) configuration ran on a laptop running MATLAB R2024b, equipped with 24.0 GB RAM and an Apple M3 CPU. The resulting mean SampEn$_{G}$ and runtimes are shown in Fig.~\ref{fig:sampen_ER_res} and Table~\ref{tab:ER_runtimes}.
\begin{table}[h]
    \centering
    \begin{tabular}{|l|c|c|c|}
    \hline
     & \multicolumn{3}{c|}{\textbf{Mean computation time (ms)}} \\
    \hline
    \textbf{Graph size} & \textbf{\(m=1\)} & \textbf{\(m=2\)} & \textbf{\(m=3\)}\\
    \hline
    \(N=30\) & 0.04   & 0.05 & 0.05  \\
    \(N=100\) & 0.19 & 0.27 & 0.37 \\
    \(N=300\) & 2.14  & 3.39 & 4.01\\
    \(N=900\) & 18.86 & 34.11 & 54.74 \\
    \(N=2700\) & 294.99 & 967.49 & 1404.48 \\
    \hline
    \end{tabular}

    \caption{Mean computation time for ER random graphs averaged across (\(m,p\)) and 20 repetitions.}
    \label{tab:ER_runtimes}
\end{table}

For $m=2$, with a random signal defined over sparse networks (small \(K\), low $p$), SampEn$_{G}$ decreases as $p$/$K$ increases, reflecting increased regularity in the system topology. However, as the graphs become denser (\(K\geq7\)), the SampEn$_{G}$ becomes less separable, continue decreasing towards low values and eventually approaches zero. This effect is more prominent for increased $m$, where the multi-hop construction substantially expand the outreach of each node's neighbourhoods. In ER graphs, this is aggravated with the random edges introducing long-range connections, which enhance reachability, leading to homogeneous patterns, consequently reducing effective variability across local patterns which SampEn$_{G}$ can exploit.

These experiments also reveal insight in the reduction of standard deviation of SampEn$_{G}$ across realisations as $N$ increases from $30$ to $2700$, indicating consistent and more stable estimates on larger graphs.

On a separate note, SampEn$_{G}$ was computationally practical in our experiments. Graphs of $N=2700$ nodes can be processed at approximately $1.4~s$ per run. For graphs with up to $N=900$, run times remain below 100$~ms$ for all embedding dimensions $m$. This is consistent with the quadratic scaling in the number of valid nodes induced by the pairwise-matching step (Sec.~\ref{par:complexityanalysis}).

This suggests that SampEn$_{G}$ is most informative on sparse to moderately-connected graphs, where multi-hop neighbourhood remains distinct.

\subsubsection{WS small-world network}
\label{Sec:WS}
Following, we investigate SampEn$_{G}$'s behaviour on small-world networks with the WS model using MATLAB's implementation \cite{MATLAB_ws}. Starting from a \(2K\)-regular ring, each edge is rewired to a randomly selected node parametrised by the number of nodes \(N\), an even lattice degree \(K\), and a rewiring probability \(\beta\), traversing between a regular ring (\(\beta =0\)) and a random graph (\(\beta =1\)). 

We consider unweighted, undirected WS graphs with $N=500$, and $K\in\{1,2,4,6\}$. On these graphs, we analyse two types of arbitrary signals: 
\begin{enumerate}

    \item Smoothed white Gaussian noise (WGN) obtained using heat-kernel diffusion (graph
    filtering) \cite{thanou_learning_2017}. The initial signal is
    \(\mathbf{n}_0 \sim \mathcal{N}(0, \mathbf{I}_N)\), and the smoothed continuous signal is:
    \begin{equation}
        \mathbf{n}_{\text{smooth}}(\tau_0) \approx e^{-\tau_0 \mathbf{L}_{\mathrm{norm}}} \mathbf{n}_0,
    \end{equation}
    where \(\mathbf{L}_{\mathrm{norm}} = \mathbf{I} - \mathbf{D}^{-1/2} \mathbf{A}\mathbf{D}^{-1/2}\)
    is the normalized graph Laplacian. The heat diffusion operation is approximated with:
    \begin{equation}
        \mathbf{n}^{(k+1)} = \mathbf{n}^{(k)} - \alpha\mathbf{L}_{\mathrm{norm}}\mathbf{n}^{(k)},
    \end{equation}
    with \(k=30\) iterations, a diffusion rate \(\tau_0 = 0.3\), and a step size \(\alpha = \tau_0 / 30\).

    \item Piecewise-constant signal constructed by partitioning the ring of \(N\) nodes into 4 sets of equal length \(K = \lfloor N/4 \rfloor\):
    \begin{equation}
        S_i =
        \begin{cases}
            \big[ (i-1) K + 1, \; i K \big], & i = 1,2,3,\\[1mm]
            \big[ 3 K + 1, \; N \big], & i = 4.
        \end{cases}
    \end{equation}
    and assigned alternating constant values of:
    \begin{equation}
        x_i =
        \begin{cases}
        +1, & i \in S_1 \cup S_3,\\
        -1, & i \in S_2 \cup S_4,
        \end{cases}
        \quad i=1,\dots,N,
    \end{equation}
    followed by additive WGN \(\mathbf{n}\sim\mathcal{N}(0,0.1^2)\). The final signal is:
    \(\mathbf{y}_{\text{piecewise-constant}}=\mathbf{x}+\mathbf{n}.
    \)
\end{enumerate}

\begin{figure}[t!]
    \centering

    \begin{subfigure}[b]{0.3\textwidth}
        \centering
        \includegraphics[width=\textwidth]{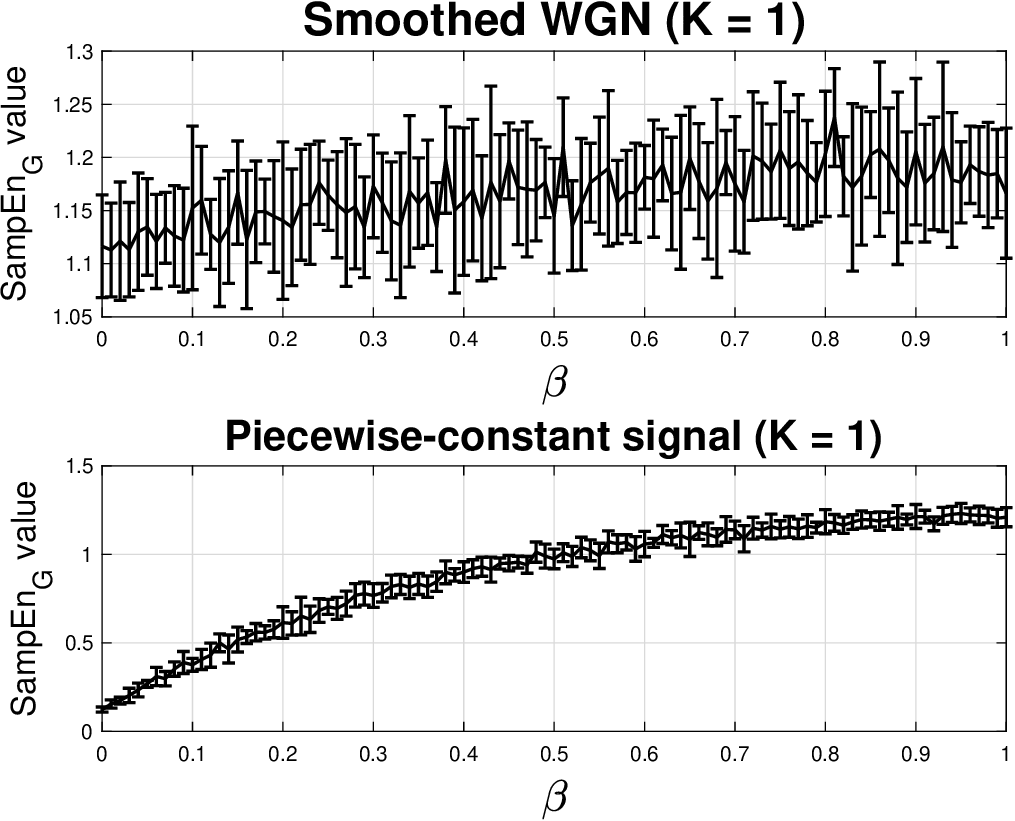}
        \caption{}
        \label{fig:WSfig1}
    \end{subfigure}
    \begin{subfigure}[b]{0.3\textwidth}
        \centering
        \includegraphics[width=\textwidth]{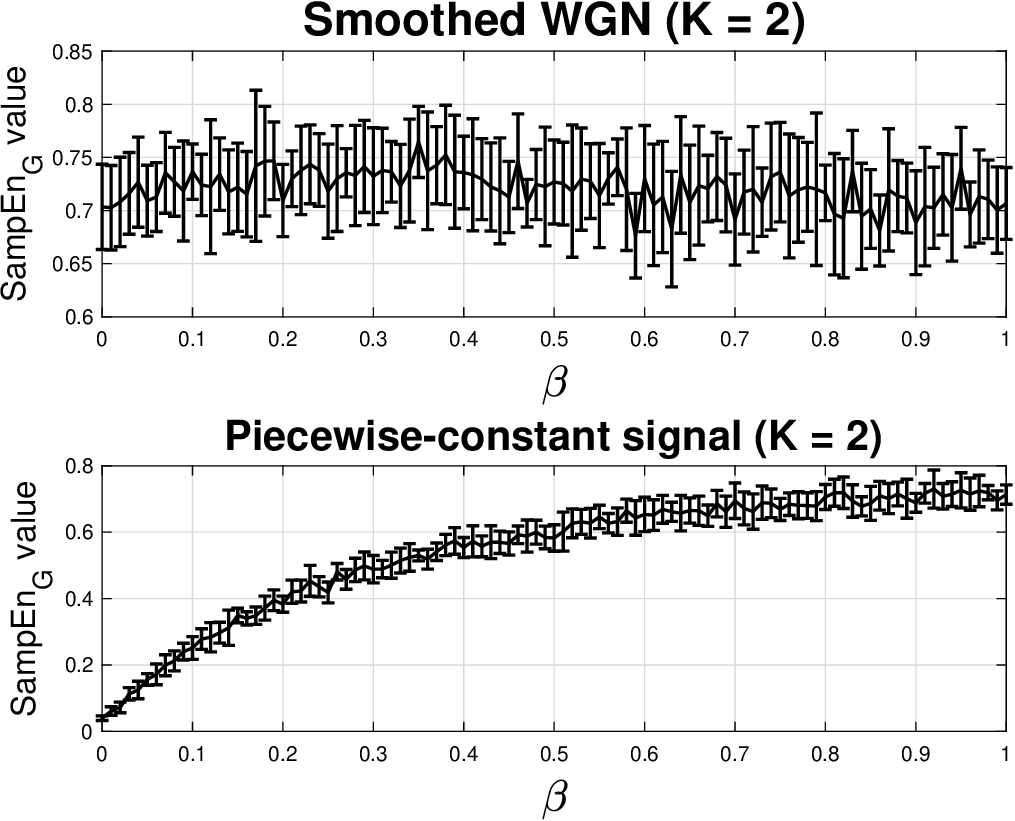}
        \caption{}
        \label{fig:WSfig2}
    \end{subfigure}
    \hfill
    \begin{subfigure}[b]{0.3\textwidth}
        \centering
        \includegraphics[width=\textwidth]{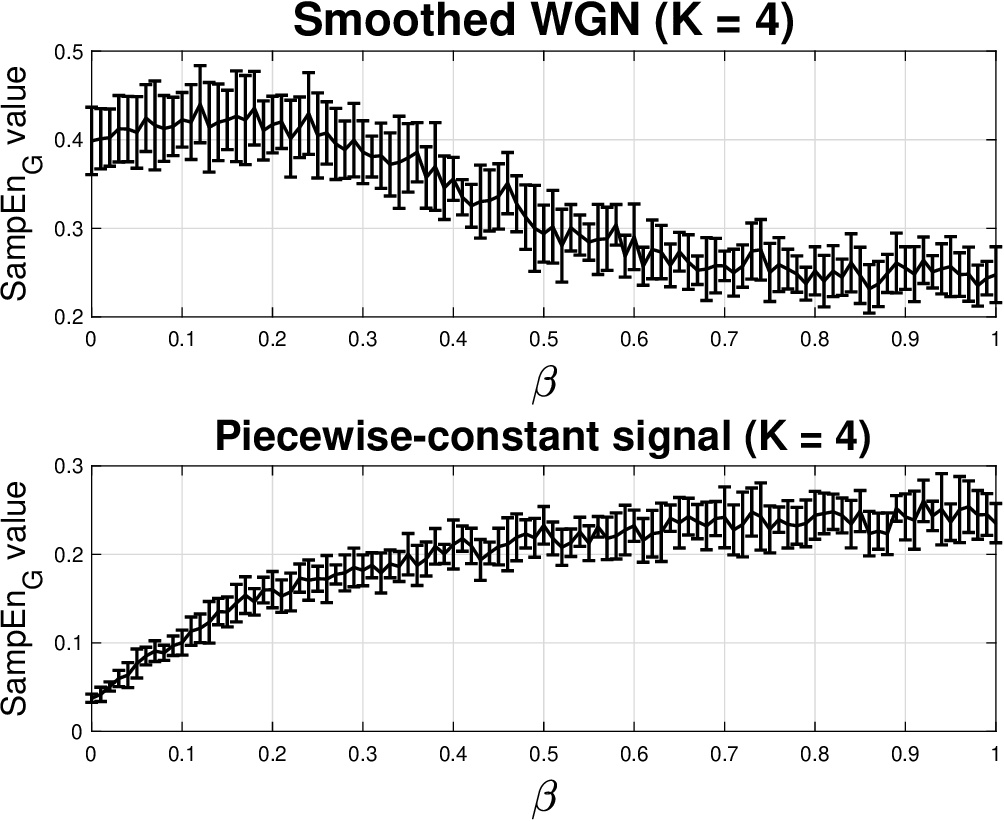}
        \caption{}
        \label{fig:WSfig3}
    \end{subfigure}
    \hfill
    \begin{subfigure}[b]{0.3\textwidth}
        \centering
        \includegraphics[width=\textwidth]{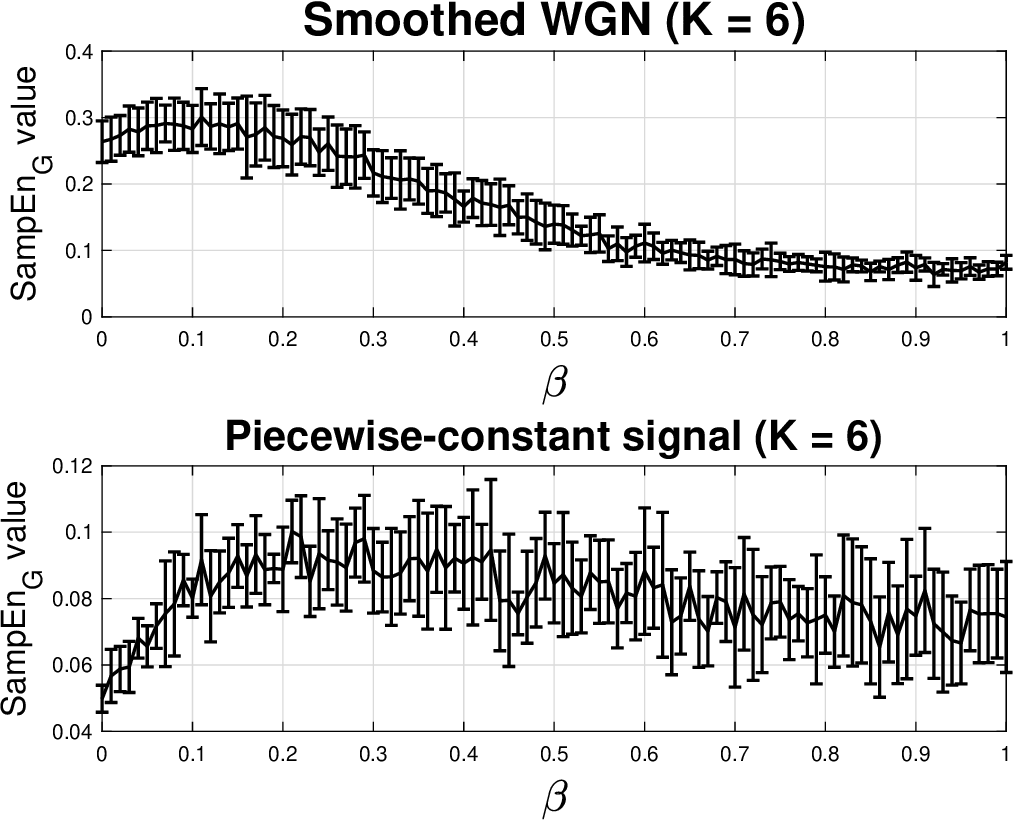}
        \caption{}
        \label{fig:WSfig4}
    \end{subfigure}

    \caption{ SampEn$_{G}$ on WS small-world networks as a function of the rewiring probability $\beta$ for lattice degrees $ 2K \in \{2,4,8,12\}$ (subfigures (a)–(d)), with $m=2$ and tolerance $r=0.2$, reporting the mean with std over 20 repetitions. Results are shown for a smoothed WGN and a piecewise-constant signal corrupted by noise.}
    \label{fig:Small-world}
\end{figure}

Figure \ref{fig:Small-world} shows the results of SampEn$_{G}$ computed with parameters \(m=2\), \(r=0.2\) as a function of \(\beta\in[0,1]\) in increment steps of $0.01$ for lattice degree \(2K=\{2,4,8,12\}\). 

As seen in Fig.~\ref{fig:WSfig1} and~\ref{fig:WSfig2}, at low degrees (\(2K=2,4\)), SampEn$_{G}$ remains approximately constant across \(\beta\) for the smoothed WGN, a random signal which local statistics are expected to not be strongly affected by the rewiring effect. On the other hand, for the piecewise-constant signal, highly regular relative to the graph topology initially, yields low SampEn$_{G}$ at \(\beta=0\). As the graph rewires (increasing \(\beta\)), SampEn$_{G}$ rises gradually to saturation, precisely reflecting how rewiring progressively misaligns the structure and the signal dynamics, and becomes effectively random to the evolving topology.

At higher degrees \(2K=8,12\) (Fig.~\ref{fig:WSfig3} and~\ref{fig:WSfig4}), the initial contrast between the two signals at low \(\beta\) remains detectable, and the difference in SampEn$_{G}$ amplitude continues to distinguish between the smoothed WGN and the piecewise-constant signal for \(\beta\in[0, 0.3]\). However, as \(\beta\) further increases, the combination of larger neighbourhoods and long-range connections amplifies the overlap in local neighbourhoods such that local structures are no longer preserved, thereby reducing disparity between matches in $m$ and $m+1$ dimensions. This reveals a limitation of the algorithm in capturing fine-grained structural differences: while long-range connections are not necessarily problematic in moderately sparse graphs; on dense graphs, they lead to an apparent increase in predictability.

\subsubsection{Kuramoto model on complex networks}\label{Sec:Kuramoto}

One of the most well-known tools for simulation of a complex system is the Kuramoto model--coupled phase oscillators comprising heterogenous natural frequencies that interact and are driven into collective synchronisation as the global coupling strength \(\lambda\) increases. Given a graph \(\mathcal{G} = (\mathcal{N}, \mathcal{E}, \mathbf{A})\), the Kuramoto dynamics on a complex topology are \cite{arenasSynchronizationComplexNetworks2008}:
\begin{equation}
\dot{\theta}_i = \omega_i + \lambda \sum_j \textbf{A}_{ij} \sin(\theta_j - \theta_i),
\qquad i = 1, \ldots, N,
\label{eq:kuramototopology}
\end{equation}
where \(\theta_i(t)\) is the phase of each oscillator \(i\), \(\omega_i\) is its natural frequency, and \(\textbf{A}_{ij}\) is the binary coupling between node \(i\) and \(j\). 

We conducted experiments on three binary, undirected topologies of \(N=300\), and a matched target mean degree of \(K\approx10\): 
\begin{enumerate}
    \item ER graph: Each pair of nodes are connected with a probability \(p\) as \eqref{eq:ERp}, resulting in a randomly-connected topology.
    \item Random Geometric Graph (RGG): We first sample \(N=300\) coordinates independently uniformly \(\textbf{x}_i\sim\mathcal{U}([0,1]^2), \quad i = 1,\ldots,N,\) and form the Euclidean distance matrix \textbf{D} where the entries are: 
    \begin{equation}
        \textbf{D}_{ij} = \lVert \textbf{x}_i - \textbf{x}_j \rVert_2.
    \end{equation}

    Each pair of node \(i\) and \(j\) is connected if their distance falls below the threshold radius \(r_c\).
    \[
	\textbf{A}_{i j}= 
	\begin{cases}
		1 & \text{if} \quad 0 < D_{ij} < r_c, \\
		0 & \text{otherwise.}
	\end{cases}
	\label{eq:AdjacencyRGG}
    \]
    We set \( r_c=\sqrt{\frac{K}{(N\pi)}}\) to target a mean degree \(K\approx10\). This yields locally clustered and spatially distributed graphs to simulate systems such as electrical grids.
    
    \item Barab\'asi-Albert (BA) graph:
    Starting from a subset of nodes \(m_0\), a new node was added with a connectivity probability \(\Pi\) to each node \(i\) within the subset depending on their degree \(k_i\) \cite{albertStatisticalMechanicsComplex2001}: 
    \begin{equation}
        \Pi(k_i) = \frac{k_i}{\sum_{j}k_j}.
    \end{equation}
    This was repeated for \(N-m_0\) steps until the network forms \(N\) nodes, we set \(m_0 = 6\) to give a resulting mean degree \(K\approx10\). The scale-free BA network is particularly useful to study social networks comprising a small number of highly-connected nodes (hubs) as opposed to the rest.

\end{enumerate}

The global synchronisation is characterised by the Kuramoto order parameter:
\begin{equation}
 r(t)\,e^{\,\mathrm{i}\psi(t)}
 = \frac{1}{N}\sum_{j=1}^{N} e^{\,\mathrm{i}\theta_j(t)},
 \label{eq:orderparam}
\end{equation}
with \(r(t)\in[0,1]\) measuring global phase coherence and \(\psi(t)\) is the instantaneous mean-field phase.

Each node \(i\) is an oscillator with a natural frequency \(\omega_i\sim\mathcal{N}(0,1)\), with an initial phase drawn from \(\theta_i(0)\sim\mathcal{U}[0,2\pi)\), for \(i=1,\ldots,N\). As phases are circular and only relative phases are meaningful, we considered the phases measured to the instantaneous mean-field phase \(\psi\), and compute SampEn$_{G}$ and DE$_{G}$ for comparison on $\cos(\theta_i - \psi)$ and $\sin(\theta_i - \psi)$.

As \(r(t)\) is not sensitive to synchronisation in local clusters ~\cite{flovikDescribingSynchronizationTopological2016}, we additionally report the mean local order \(L_{loc}\) to measure the local neighbourhood coherence, independent of the global alignment \cite{flovikDescribingSynchronizationTopological2016}:
\begin{equation}
  \beta_i =
\frac{1}{k_i} \left| \sum_{j} A_{ij}\, e^{\mathrm{i}\theta_j} \right|,
  \label{eq:localorderb}
\end{equation}
where \(k\) is the degree of each node. We then summarised by the network level for a local synchronisation measure:
\begin{equation}
  L_{loc} = \frac{1}{N}\sum_{i=1}^{N}
  \beta_i.
  \label{eq:localorder}
\end{equation}

\begin{figure}
    \centering
\begin{subfigure}{\linewidth}
    \includegraphics[width=\linewidth]{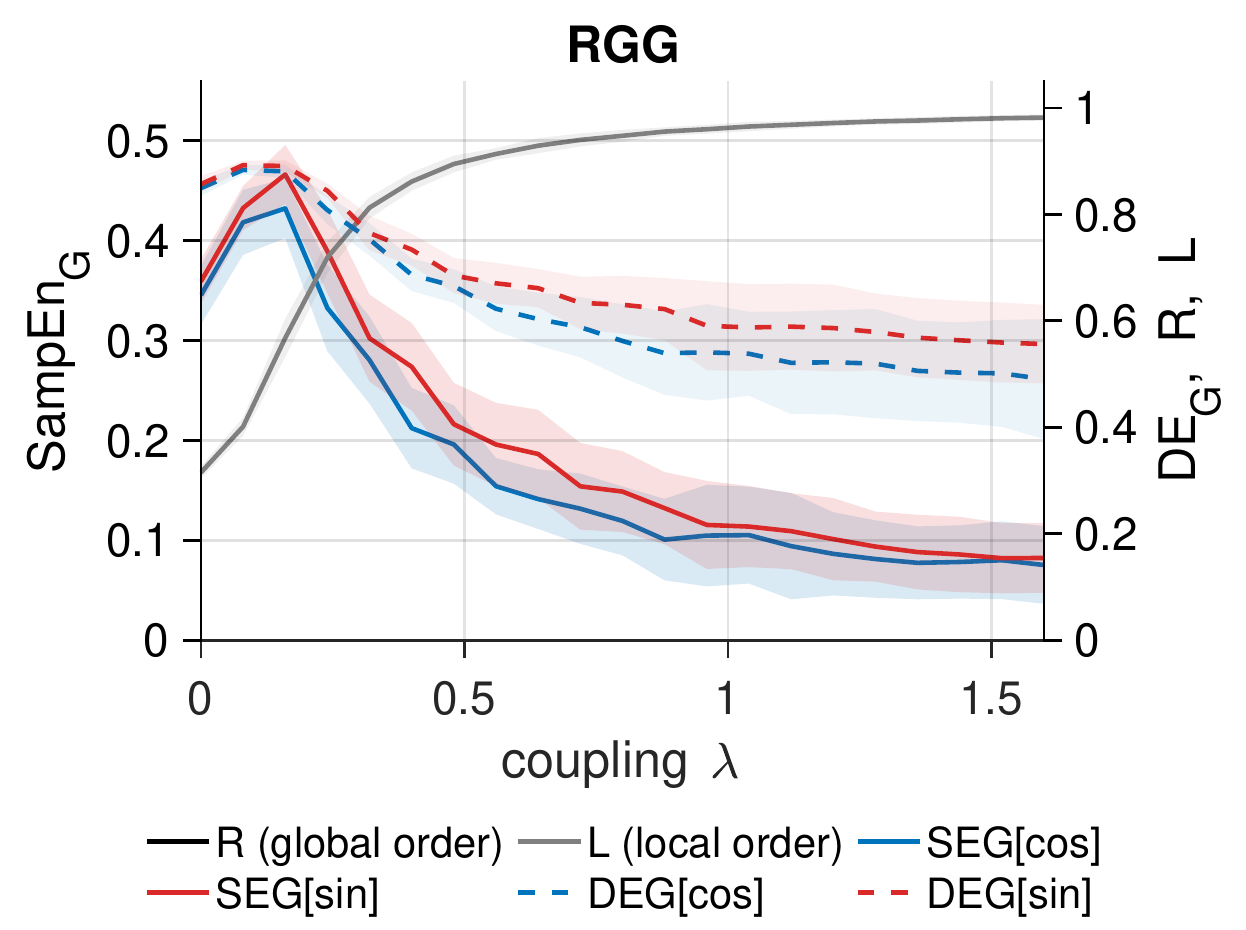}
    \caption{Random geometric graph (RGG)}
    \label{fig:KMRGG}
\end{subfigure}\hfill
\begin{subfigure}{\linewidth}
    \includegraphics[width=\linewidth]{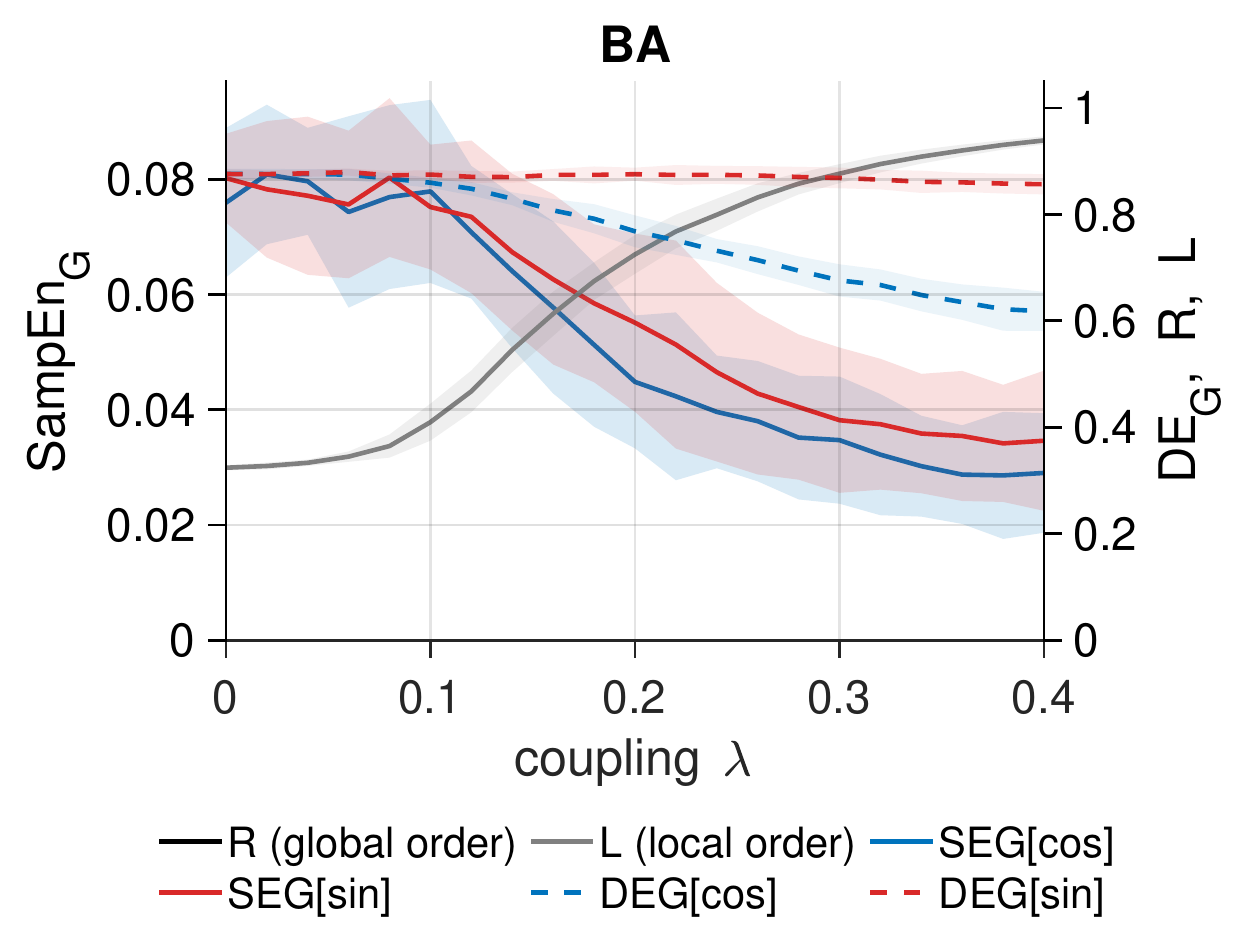}
    \caption{Barab\'asi-Albert (BA)} 
    \label{fig:KMBA}
\end{subfigure}\hfill
\begin{subfigure}{\linewidth}
    \includegraphics[width=\linewidth]{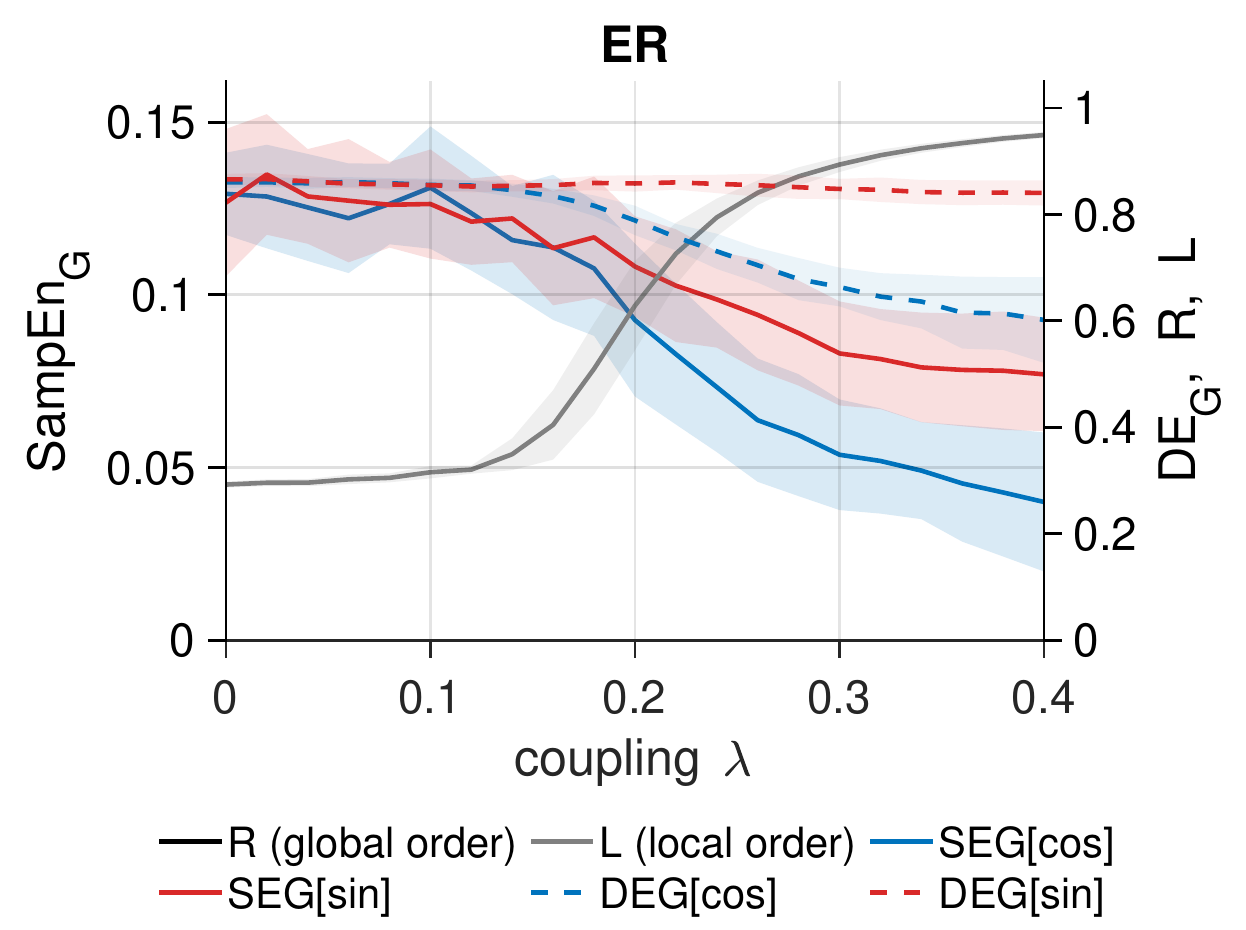}
    \caption{Erd\H{o}s-R\'enyi (ER)}
    \label{fig:KMER}
\end{subfigure}
    \caption{SampEn$_G$ and DE$_G$ on the Kuramoto model versus coupling $\lambda$ for (a) RGG, (b) BA, (c) ER topologies ($N=300$, $K\approx10$, $m=2$, $r=0.2$, mean $\pm$ std over 20 runs). Right axis: global order $R$ and local order $L_{loc}$. SampEn$_G$ peaks on the RGG where local synchronisation precedes global.}
\label{fig:KMmodel}
\end{figure}

 For each topology, we integrated with a fourth-order Runge-Kutta scheme with \(\Delta t =0.05\). We discarded a transient of $30$ time units, within a window we averaged with five equally spaced snapshots taken at each coupling. This coupling \(\lambda\) was swept for a range of values over which \(R\) goes to full synchronisation. At each \(\lambda\), we computed SampEn$_{G}$ ($m=2,r=0.2)$, and DE$_G$ ($m=2,L=1,c=4$) over 20 independent realisations and report their mean and standard deviation for RGG, BA, and ER graphs in Fig.~\ref{fig:KMmodel}, with the global order parameter \(R\) (the time-average of \(r(t)\)), and the mean local order parameter \(L_{loc}\).

As can be seen, all three experiments yield consistent behaviour with cosine and sine signals on each topology: SampEn$_{G}$ decreases as coupling \(\lambda\) increases, indicating higher network regularity. This confirms that SampEn$_{G}$ response reflects the underlying system organisation.  

On the RGG experiment, SampEn$_{G}$ is non-monotonic as coupling \(\lambda\) increases, the initial arch coincides with a clear separation of \(L_{loc}\) and \(R\), indicating locking of oscillators with their immediate neighbours forming local synchronised patches before global synchronisation. This maximises the diversity of local multi-hop patterns, driving SampEn$_{G}$ to a peak. 

This initial peak of SampEn$_{G}$ exceeds its across-realisation standard deviation by a factor of \(2.25\) (cosine) and \(2.83\) (sine), confirming the bend is a reproducible feature, whereas the corresponding DE$_{G}$ rise lies within its own variability (\(0.55\) and \(0.70\), see Table~\ref{Table:KMbend}). However, this observed behaviour is topology-dependent. On the ER and BA graphs, which connectivity is not spatially embedded, locally synchronised patches were not formed: \(L_{loc}\) and \(R\) rise together through an exponential transition. SampEn$_{G}$ shows a monotonic decrease accordingly, demonstrating that SampEn$_{G}$ responds not only to the degree of synchronisation, but also to the local synchronisation, distinguishing a topology from the others.

The two measures also differ markedly in their sensitivity to the choice of phase projection (see Table~\ref{Table:KMsd}). On the ER and BA networks, DE$_G$ computed on the sine projection is primarily flat, varying by only \(3.0\%\) and \(2.6\%\) of its baseline across the range of \(\lambda\), as the networks transition from chaos to global synchronisation. DE$_{G}[cos]$ over the same topologies varies by approximately $27\%$. In contrast, SampEn$_{G}$ responds consistently on both projections (relative ranges of \(45.7\%\) and \(57.5\%\) on the ER and BA networks) over the transition regardless of the projection function. However, this projection-dependent blindness of DE$_{G}$ is confined to the dense, non-local topologies--on RGG both DE$_{G}$ projections vary at \(46.3\%\) (cosine) and \(39.2\%\) (sine).


We note that the observed contrast between the two projections may be attributed to the binned-symbol frequencies of the sine signal forming a broad distribution that remains nearly unchanged across coupling, as a result, DE$_{G}[sin]$ remains high and approximately constant. On the other hand, SampEn$_{G}$, which computes an estimate of pattern recurrence via a continuous threshold \(\epsilon=r\times SD\), recognises the reorganisation. As the relationship between each node's value and its neighbourhood's value become more correlated, this shifts the matching ratio of $m$ and $m+1$, in which SampEn$_{G}$ is sensitive to. SampEn$_{G}$ avoids the blindness in DE$_{G}$ as it measures the conditional entropy of multi-hop neighbourhoods directly in the continuous space.

\begin{table}[t]
\centering
\setlength{\tabcolsep}{3pt}
\caption{Robustness of the initial peak in the entropy curves, measured as the standard deviation across-realisation \(\sigma\), bend-to-noise ratio (BNR) $= [\max_\lambda\overline{H}-\overline{H}(\lambda_{\min})]/\overline{\sigma}$ over $20$ realisations. $\mathrm{BNR}>1$ indicates the interior rise exceeds the across-realisation noise; $\mathrm{BNR}<1$ indicates it is within noise. A pronounced, robust bend appears only for SampEn$_{G}$ on the RGG.}
\label{tab:supp_bnr}
\begin{tabular}{lcccc}
\toprule
Topology & SampEn$_{G}$[cos] & SampEn$_{G}$[sin] & DE$_G$[cos] & DE$_G$[sin] \\
\midrule
RGG & 2.25 & 2.83 & 0.55 & 0.70 \\
ER & 0.11 & 0.52 & 0.02 & 0.03 \\
BA & 0.41 & 0.01 & 0.00 & 0.27 \\
\bottomrule
\end{tabular}
\label{Table:KMbend}
\end{table}

\begin{table}[t]
\centering
\setlength{\tabcolsep}{3pt}
\caption{Relative range of the Kuramoto entropy curves, measured by the percentage variation of the mean curve across the coupling sweep over \(\lambda\). A small value indicates the measure is essentially flat across the sweep, i.e.\ unresponsive to the synchronisation transition. DE$_G$[sin] is unresponsive on the ER and BA networks, whereas SampEn$_{G}$ responds on both projections.}
\label{tab:supp_pctvar}
\begin{tabular}{lcccc}
\toprule
Topology & SampEn$_{G}$[cos] & SampEn$_{G}$[sin] & DE$_G$[cos] & DE$_G$[sin] \\
\midrule
RGG & 103.4\% & 107.0\% & 46.3\% & 39.2\% \\
ER & 70.4\% & 45.7\% & 30.1\% & 3.0\% \\
BA & 68.8\% & 57.5\% & 29.5\% & 2.6\% \\
\bottomrule
\end{tabular}
\label{Table:KMsd}
\end{table}

\subsection{Real-world graph signals}
In this section, we evaluate SampEn$_G$ on four real-world graph signal datasets to assess its performance in practical settings with evaluation on its robustness to parameters in graph construction such as kernel settings and edge perturbations.

\subsubsection{Weather station dataset}\label{Sec:weatherstation}
We considered temperature data collected from 37 distributed weather stations in Brittany during January 2014 \cite{giraultStationaryGraphSignals2015}. Following \cite{fabila-carrascoPEG2022}, the weather stations were modelled as a weighted, undirected graph where the edges are the pairwise Euclidean distances between stations computed through a Gaussian kernel, encoding the spatial distribution as:
\begin{equation}
    \textbf{W}_{ij} = \begin{cases}
        \exp\!\left( \frac{-d(i,j)^2}{2\sigma_1^2}\right), & \text{if } d(i,j)\leq \sigma_2 \\
        0, & \text{otherwise.}
        
    \end{cases}
    \label{eq:weatherstationadj}
\end{equation}
where \(\sigma_1^2=5.1\times10^8\) is the kernel scale, and \(\sigma_2=10^5\) is the distance threshold.

We conducted the analysis on an ensemble of temperature readings at 4:00 (night) and 14:00 (day) over 31 days. We computed SampEn$_{G}$ as a function of tolerance $r$, and report the mean and standard deviation across the 31 observations.

\begin{figure}[h!]
    \centering
    \includegraphics[width=0.9\linewidth]{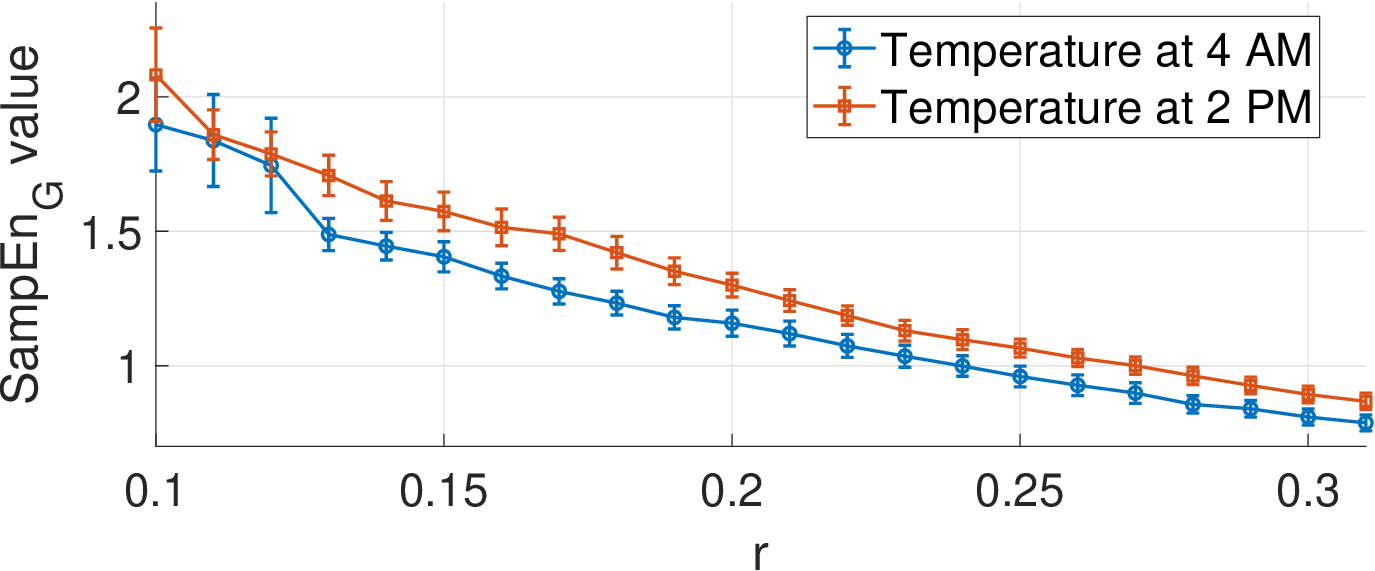}
    \caption{Averaged SampEn$_{G}$ of 31 temperature observations at 4:00 and 14:00, with respect to $r$ ($m=1$).}
    \label{fig:temp_res}
\end{figure}

As shown in Fig.~\ref{fig:temp_res}, daytime (2PM) consistently yields higher mean SampEn$_G$ entropy values than nighttime (4AM) across a wide range of $r$ values. The findings align with physical expectations: daytime temperatures exhibit more complex patterns due to interactions among solar radiation, atmospheric, and geographic dynamics; while nighttime temperatures tend to exhibit more stable and regular patterns in the absence of solar activity. These results demonstrate the capability of SampEn$_{G}$ to discriminate between different dynamical regimes on a real spatially distributed weather-station network. It is noteworthy that the range where the two curves are well-separated is $0.15-0.22$, coherent with typical choices in classical SampEn \cite{sampen_typical_param}.

\subsubsection{Robustness to graph construction}\label{Sec:robustness_construction}

The weather station connectivity in Sec.~\ref{Sec:weatherstation} is a similarity graph derived from inter-station geometry, which parameters may be any reasonable value of choice. In this subsection, we assess the sensitivity of SampEn$_{G}$ to the kernel parameters in graph construction~\eqref{eq:weatherstationadj} using the weather station experiment: kernel scale \(\sigma_1\) controls the kernel weight decay thus weight distribution; while the distance threshold \(\sigma_2\) controls the network density.  

We swept each parameter over four
orders of magnitude with the other fixed at its reference value from \cite{fabila-carrascoPEG2022},
and similarly report SampEn$_G$ at the two times of day (2~PM and 4~AM), the day--night difference, and graph connectivity (normalised Fiedler eigenvalue \(\frac{\lambda_2}{\text{max}(\lambda)}\) and mean edge weight) alongside for comparison as a function of the kernel scale.

\begin{figure}
    \centering
    \begin{subfigure}{.9\linewidth}
        \centering
        \includegraphics[width=\linewidth]{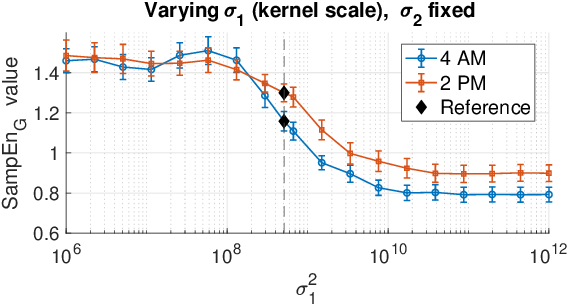}
        \caption{SampEn$_G$ at 2\,PM and 4\,AM}
        \label{fig:s1_sampen}
    \end{subfigure}
    \hfill
    \begin{subfigure}{\linewidth}
        \centering
        \includegraphics[width=.9\linewidth]{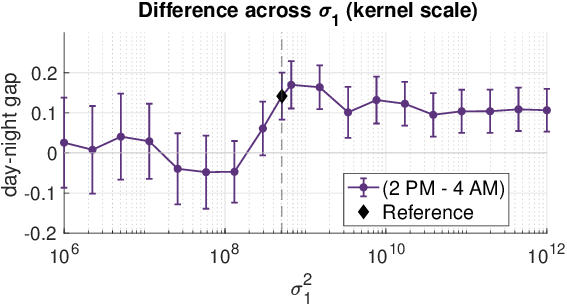}
        \caption{Day--night gap}
        \label{fig:s1_gap}
    \end{subfigure}
    \hfill
    \begin{subfigure}{\linewidth}
        \centering
        \includegraphics[width=.9\linewidth]{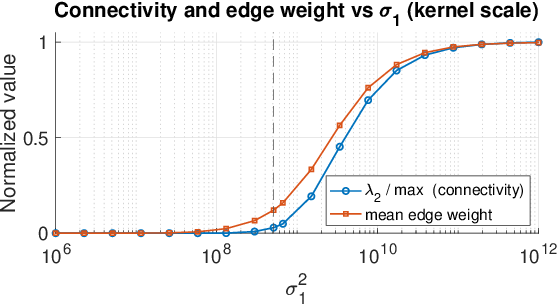}
        \caption{Connectivity and edge weight}
        \label{fig:s1_connectivity}
    \end{subfigure}
    \caption{Sensitivity of SampEn$_G$ to the kernel scale $\sigma_1^2$ (with $\sigma_2$ fixed) on the weather-station graph. (a) SampEn$_G$ at 2\,PM and 4\,AM; (b) the day--night gap; (c) normalised graph connectivity (Fiedler eigenvalue) and mean edge weight. The day--night discrimination stays positive across the sweep, breaking down only at small $\sigma_1^2$ where the graph weights \(\approx0\) (essentially disconnected) and at large $\sigma_1^2$ where edge weights homogenise toward unity (saturation regime). The reference value of \(\sigma_1^2=5.1\times10^8\) is marked.}
    \label{fig:sigma1vary}
\end{figure}

Fig.~\ref{fig:sigma1vary} shows the results of SampEn$_{G}$ across the sweep of \(\sigma_1^2\) (kernel scale). At small $\sigma_1^2\leq 3\times 10^8
$, the kernel decays steeply, resulting in
most weights becoming negligible--reflected by normalised Fiedler
eigenvalue collapsing towards zero in Fig.~\ref{fig:s1_connectivity}--and the graph becomes effectively disconnected spectrally. Thus, the entropy estimates becomes unreliable. As \(\sigma_1^2\) increases past the reference value, SampEn$_{G}$ decreases at both times of the day. The observation is consistent with the saturation regime described in Sec.~\ref{fig:MIX_res}: as the kernel drives edge weights homogeneity towards unity, walk-weighted multi-hop averages concentrate around the graph signal's mean, reducing the differences in patterns  for the algorithm. Nevertheless, SampEn$_{G}$ displays a consistent positive day--night difference even as the graph approaches full connection (both connectivity
and mean edge weight rising toward unity).

\begin{figure}
    \centering
    \begin{subfigure}{0.9\linewidth}
        \centering
        \includegraphics[width=\linewidth]{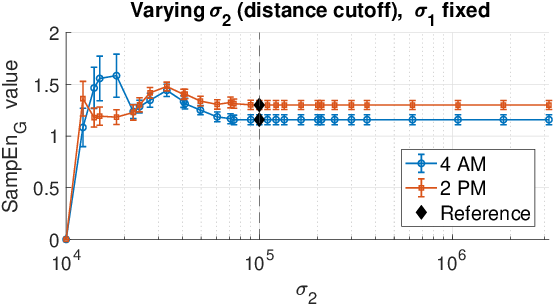}
        \caption{}
        \label{fig:s2_sampen}
    \end{subfigure}
    \hfill
    \begin{subfigure}{0.9\linewidth}
        \centering
        \includegraphics[width=\linewidth]{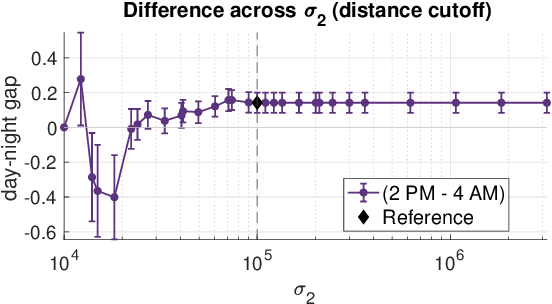}
        \caption{Day--night gap}
        \label{fig:s2_gap}
    \end{subfigure}
    \caption{Sensitivity of SampEn$_G$ to the distance cutoff $\sigma_2$ (with $\sigma_1$ fixed) on the weather-station graph. (a) SampEn$_G$ at 2\,PM and 4\,AM; (b) the day--night gap. Below $\sigma_2\approx4\times10^{4}$ the graph is too sparse for reliable estimates and the gap fluctuates around zero; above it both curves stabilise at a positive day--night difference. The reference value of \(\sigma_2=10^5\) is marked.}
    \label{fig:sigma2vary}
\end{figure}

Fig.~\ref{fig:sigma2vary} shows the sweep across the distance cutoff \(\sigma_2\). At small cutoff \(\sigma_2\leq 4\times10^4\), the graphs are too sparse for reliable estimates, where the day--night difference is fluctuant and centred around zero. Exceeding \(\sigma_2 =4\times10^4\), both curves stabilise and settle at a positive day--night difference for larger cut-offs, maintaining the discriminative behaviour.

Overall, the sensitivity analysis indicates the day--night discrimination achieved by SampEn$_{G}$ is robust to variations in the kernel parameters. The breakdown occurred only at extreme parameter choices leading to the graph becoming disconnected.

\subsubsection{Robustness to edge perturbation}\label{Sec:perturbedge}

We further assessed the robustness of SampEn$_{G}$ to edge perturbation, we evaluated the day--night discrimination under controlled random perturbation from the reference graph from Sec.~\ref{Sec:weatherstation}. The reference graph is undirected, with a density of \(0.542\) (\(|\mathcal{N}|=37\), \(|\mathcal{E}| = 361\)), and mean edge weight \(0.12\). 

We considered two perturbation types and generated results from 20 realisations each.
\begin{enumerate}
    \item Edge deletion removes a percentage \(\%\) of the existing edges uniformly at random without replacement.
    \item Edge addition introduces a percentage \(\%\) of new edges between disconnected pairs, selected uniformly at random. Each new edge was assigned the mean edge weight from the reference graph, rather than using the Gaussian kernel value from geometric distance, as it would carry almost near-zero weights at long range, to ensure the perturbation is substantive.
\end{enumerate}

The day--night difference (2~PM-4~AM) in SampEn$_{G}$ ($m=1$, $r=0.2$) was computed on each perturbed graph in steps of \(10\%\),
and we report the mean and standard deviation across realisations for each perturbation type.

\begin{figure}[t]
    \centering
    \includegraphics[width=.9\linewidth]{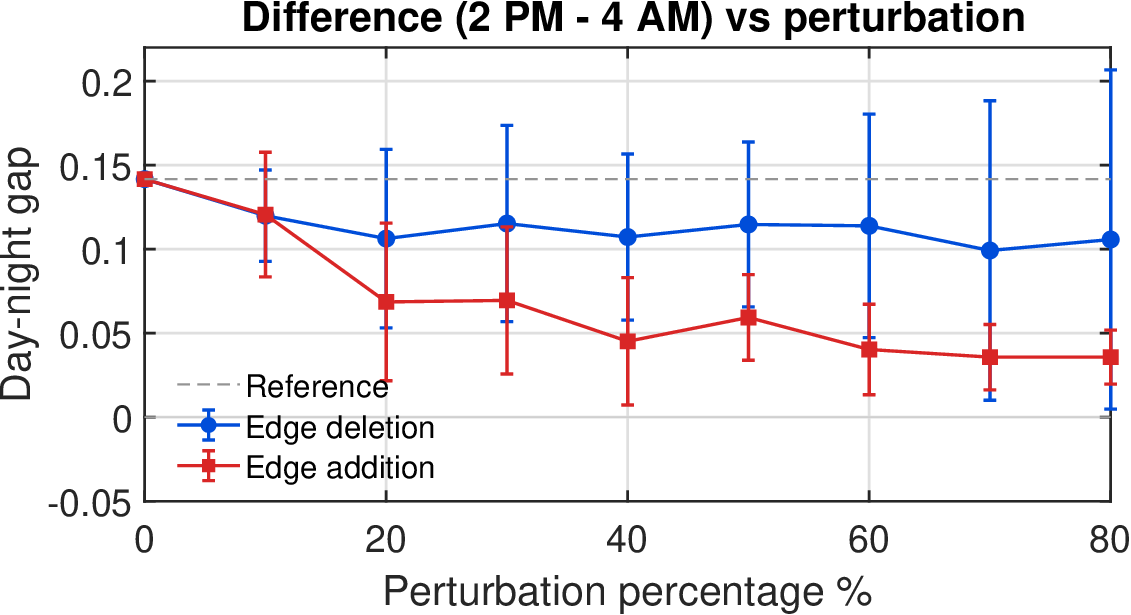}
    \caption{Day--night gap (2\,PM$-$4\,AM) of SampEn$_G$ under random edge deletion and addition on the weather-station graph ($m=1$, $r=0.2$, mean $\pm$ std over 20 realisations); dashed line is the reference. The gap stays positive throughout; addition degrades it faster by breaking spatial locality.}
    \label{fig:perturbgap}
\end{figure}

As seen in Fig.~\ref{fig:perturbgap}, both perturbations retains a positive day--night difference throughout. Under edge deletion, the gap stayed relatively close to the reference value across the entire range, with larger variance at high perturbation percentage \(\%\) as the networks become progressively sparser. 

In contrast, edge addition, on top of a half-connected graph (density $0.542$) produces a reduction in the day--night gap beyond perturbation percentage $20\%$. This is consistent with the experimental design: new edges are assigned the mean reference weight rather than the near-zero kernel value, introducing long-range connections simultaneously. As a result, they destroy the spatial locality in topology, greatly increasing the outreach thus overlap on the local weighted neighbourhood structure on which SampEn$_{G}$ depends.

As the graph approaches full connection, this drives the saturation regime, mirroring the behaviour observed in Fig.~\ref{fig:sigma1vary} at a large kernel-scale threshold. Despite this strong intervention, the day--night gap remains positive at every tested perturbation level.

\subsubsection{Intel Berkeley lab sensor network}
We next assessed the algorithm on a \textit{directed}, weighted wireless sensor network dataset collected at the Intel Berkeley Research lab \cite{intelLabData2004}. We focused on a four-day window (March 19th, 5:00 AM to March 23rd, 5:00 AM), during which all sensors exhibit inter-sample intervals under 40 minutes.

We retained a subset of 23 sensors that had at least 60\% the maximum number of light-intensity observations across sensors. The data were then re-sampled at 5-minute intervals, with missing values interpolated linearly. Following, we extracted daytime (8:00 AM to 5:00 PM) and nighttime (8:00 PM to 5:00 AM) segments, each spanning 9 hours (108 samples), to contrast system dynamics between active and inactive periods based on working hours. The adjacency matrix was populated as:
\[ \footnotesize
	\mathbf{W}_{i j}= 
	\begin{cases}
		pr(i,j) & \text{if there exist a connection from } i \text{ to } j,  \\
		0 & \text{otherwise.}
	\end{cases}
\]
where $pr(i,j)$ denotes the probability of a successful message transmission from node $i$ to $j$ as provided in the dataset. Using this, we computed SampEn$_{G}$ on each time point across four day windows with $m\in\{1,2\}$, reporting the mean and standard deviation, varying $r$ for daytime vs nighttime in a similar fashion as the weather station experiment.

\begin{figure}[t!]
    \begin{subfigure}[b]{\linewidth}
        \centering
        \includegraphics[width=0.9\textwidth]{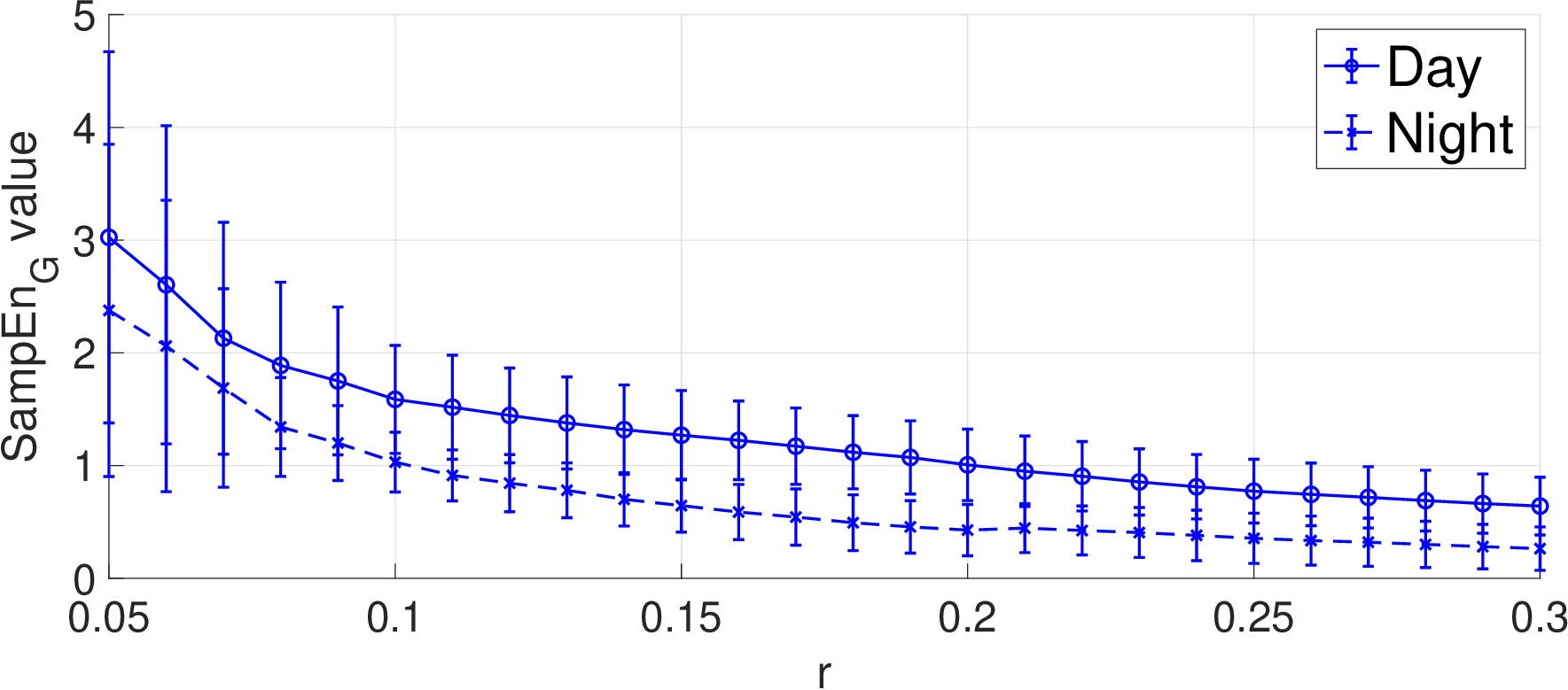}
        \caption{}
        \label{fig:intelfig1}
    \end{subfigure}
    \begin{subfigure}[b]{\linewidth}
        \centering
        \includegraphics[width=.9\textwidth]{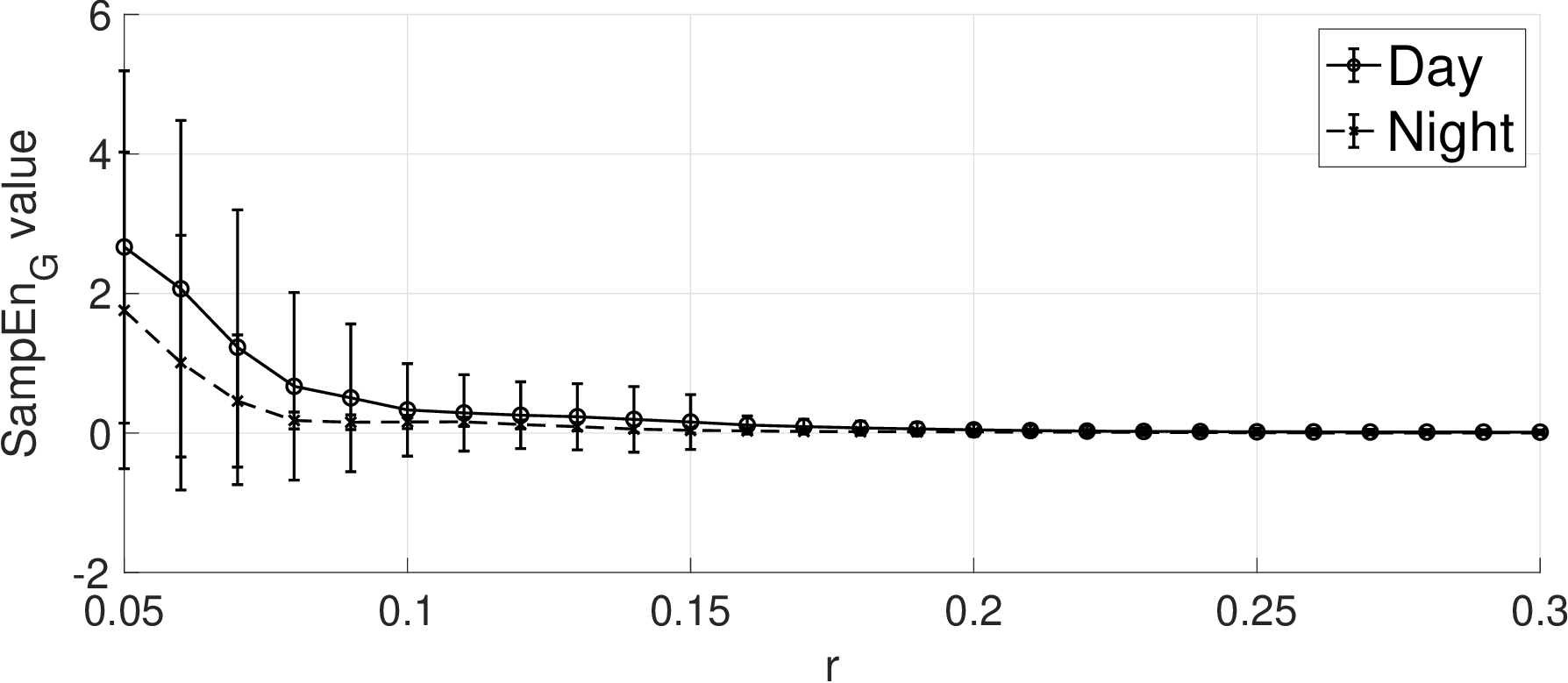}
        \caption{}
        \label{fig:intelfig2}
    \end{subfigure}
    \caption{SampEn$_G$ for the Intel Berkeley lab sensor network ($N=23$) as a function of the tolerance $r$, for daytime and nighttime light-intensity data: (a) $m=1$, (b) $m=2$. Daytime values are consistently higher, with the clearest separation near $r\in[0.16,0.2]$.}
    \label{fig:Intel-lab}
\end{figure}

As seen Fig.~\ref{fig:intelfig1}, the daytime entropies were readily differentiable from the nighttime entropy value for $m=1$ across \(r\) with a well-separated gap between the two curves at \(r\in[0.16,0.2]\), consistent with existing literature at \(r\approx0.2\) \cite{sampen_typical_param}. Similarly for $m=2$ in Fig.~\ref{fig:intelfig2}, the SampEn$_{G}$ values were consistently higher for daytime than those of the nighttime. 

This behaviour aligns with our expectations: greater dynamics and light variations thus higher SampEn$_{G}$ value in daytime attributed to solar activities, human interaction and interference within the lab during office hours; while light intensity at nighttime is often more stable thus predictable due to the absence of the mentioned activities, constituting lower SampEn$_{G}$ value. It is noteworthy that the experiment was conducted on a graph of size \(N=23\), with short data lengths (108 data points per window).

The overlap in the two curves in Fig.~\ref{fig:intelfig2} is likely attributed to the small graph size, nonetheless, results illustrated substantiates the algorithm's robustness and performance to short data and graph topology representing wireless connectivity between sensors, other than geometric distance as the previous experiment. 

\subsubsection{Freeway traffic (FT-AED)}\label{Sec:ft-aed}
We performed analysis on a complex, dynamical system on a networked traffic flow using the FT-AED dataset collected from 196 sensors (4 lanes at 49 mile-markers) along a freeway towards Nashville, Tennessee, USA \cite{coursey2024ftaed}. The dataset consists 3,763,200 speed measurements recorded every 30 seconds on weekday mornings between 4:00 AM and 12:00 PM during October, 2023. 

Over the time interval, the traffic-flow time series displays dynamical phase transitions: an initial smooth-flow phase, a subsequent congested phase, and late-hour phase where occasional anomalous events occur. We analysed the nonlinear dynamics and bifurcation of the structural traffic network in this section. 

We represent the 196 sensors as nodes arranged by lane and increasing mile-marker order, and construct a binary adjacency matrix. The topology is designed to capture (i) longitudinal coupling along each lane (propagation of traffic) and (ii) lateral coupling between neighbouring lanes (lane-changing effects). Within-lane connectivity is kept undirected in both graph variants, while directionality is introduced only in the cross-lane connections to encode a downstream flow constraint.

\begin{figure}[t!]
    \centering
    \includegraphics[width=\linewidth]{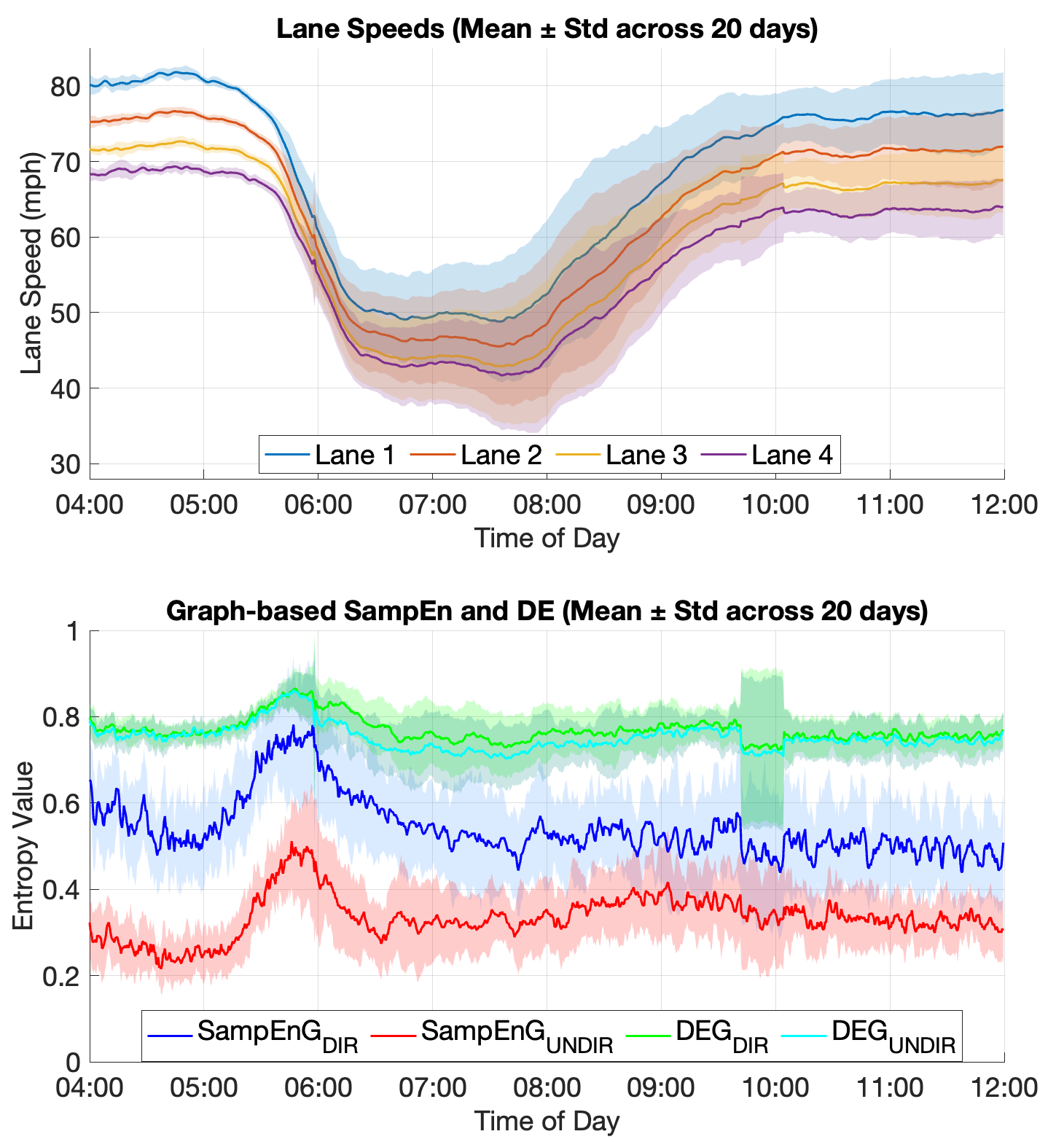}
    \caption{Traffic-flow analysis on the FT-AED dataset (mean $\pm$ std over 20 weekday mornings). Top: lane speeds. Bottom: entropy measures --- directed SampEn$_G$ (blue), undirected SampEn$_G$ (red), directed DE$_G$ (green), undirected DE$_G$ (cyan). Directed SampEn$_G$ peaks $\approx20$~mins before the other measures at congestion onset.}
    \label{fig:EntropyFT-AED}
\end{figure}

Formally, let $(l,k)$ index the sensor at lane $l\in\{1,2,3,4\}$ and mile-marker position $k\in\{1,\dots,49\}$ (in increasing downstream order). For within-lane coupling, we add edges between adjacent sensors $(l,k)$ and $(l,k+1)$ for all valid $k$. For cross-lane coupling, each sensor $(l,k)$ is connected to downstream sensors in lane $l+1$. In the directed graph we set $A_{(l,k),(l+1,k')}=1$ for all $k'\geq k$, while the undirected variant uses the corresponding symmetrised edges between the same node pairs. An illustration of the resulting networks is shown in Supplementary materials, Fig.~S3-S4. The experimental results, along with the mean traffic speed over 20 weekday mornings are presented in Fig.~\ref{fig:EntropyFT-AED}.

Notably, SampEn$_{G}^{Dir}$ attains its peak at 05:30, about 20 minutes earlier than the peaks of SampEn$_{G}^{Undir}$ and both DE$_{G}$ variants (05:50). This suggests that the directed topology, encoding traffic upstream/downstream influence, together with the conditional-entropy formulation of SampEn$_{G}$, provides an early-warning signal on the commencing congestion. 

After these peaks, all entropy measures decline as the traffic proceeds slowly after 06:00. At 08:00, the traffic regains its motion and shows increased variability later in the morning. All measures register a distinct change in the 09:45-10:00 window, consistent with the tail of the rush hour. The subsequent rise appears associated with an anomaly-rich period from 10:00 and 12:00 across the 20 weekdays. SampEn$_{G}^{Undir}$ is accompanied with an increase in entropy and decrease, but rather subtle as compared to the other measures. 

Overall, SampEn$_{G}$ displays a larger-amplitude rise at congestion onsets and larger declines as traffic recovers, demonstrating greater sensitivity to traffic-state transitions compared to DE$_{G}$. When computed on the directed topology, the conditional nature of SampEn$_{G}$, compared to DE$_{G}$ capturing local ordinal complexity, makes it better suited for detecting transitions and providing early prediction of traffic congestion.

\subsubsection{PeMS dataset}\label{Sec:PeMS}
Lastly, we extended the traffic network analysis to a large-scale traffic dataset--California Performance Measurement System (PeMS) 5-minute station readings dataset \cite{choeFreewayPerformanceMeasurement2001} consisting \(18{,}741\) sensors in all districts \(\in\{3,4,5,6,7,8,10,11,12\}\), across 61 weekdays (1 January to 31 March) in 2019 excluding American bank holidays.

The adjacency matrix was constructed with a thresholded Gaussian kernel as \eqref{eq:weatherstationadj}, where \(d(i,j)\) is pairwise geodesic sensor distances, keeping connections within a radius (\(\rho=4\)~km), kernel scale set to the standard deviation of all pairwise distances here, and pruning weights below a threshold of \(0.01\), following the construction used in \cite{liuLargeSTBenchmarkDataset2023}. Of the \(18{,}741\) sensors, \(565\) were fully isolated in the adjacency (no neighbours within \(\rho\)). These were excluded, leaving \(18{,}176\) connected nodes in the computation of SampEn\(_{G}\) and DE\(_{G}\).

The speed field follows the PeMS convention where detectors without valid speed report 0. In the same way in \cite{liuLargeSTBenchmarkDataset2023}, the PeMS 5-minute station readings for 2019 were aggregated to 15-minute resolution, we extracted the flow-weighted mean speed, and missing values are linear interpolated across time. 

\begin{figure}[t]
    \centering
    \includegraphics[width=\linewidth]{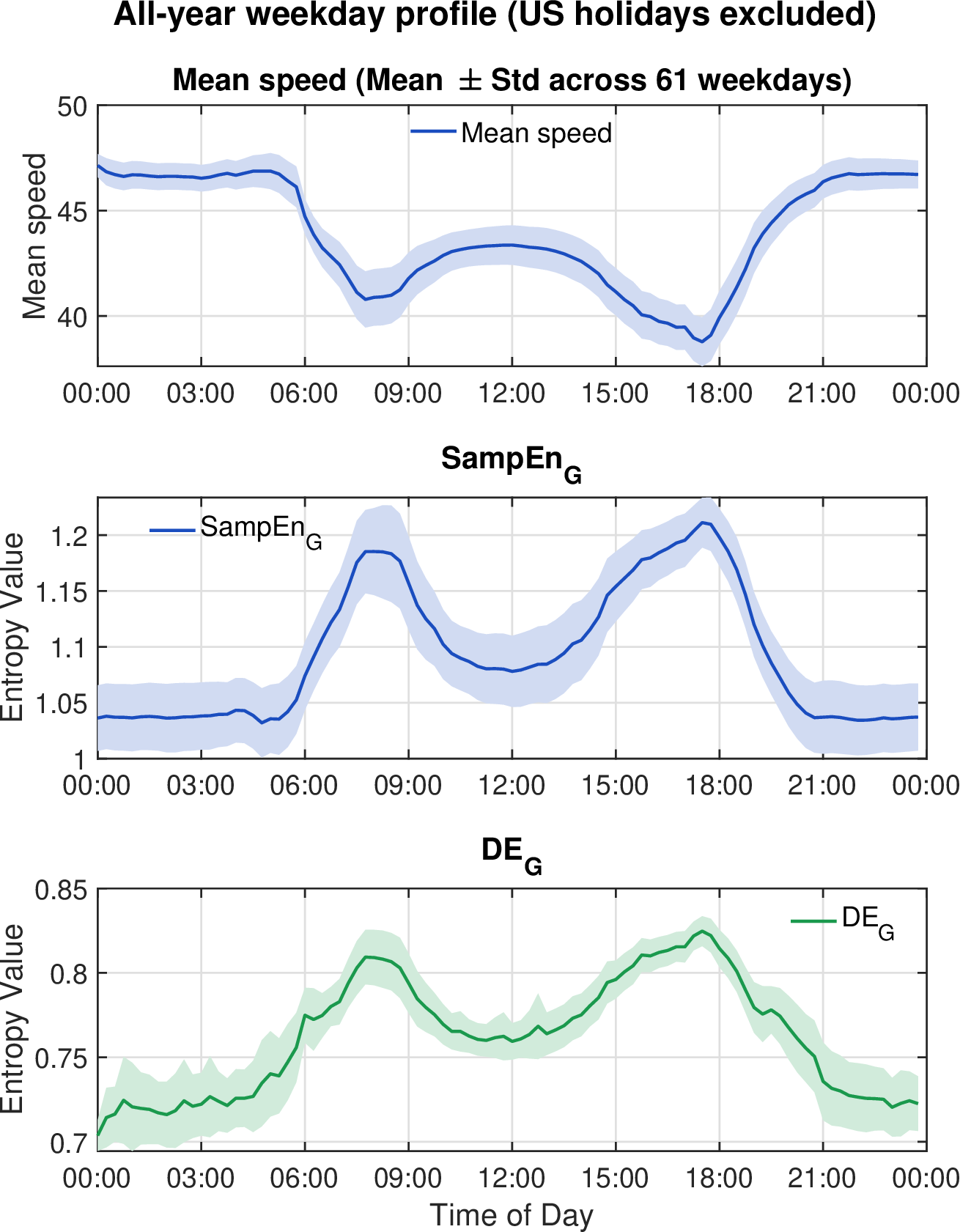}
    \caption{61-weekday profile of the PeMS network ($N=18,176$, holidays excluded): mean speed, SampEn$_G$, and DE$_G$ (mean $\pm$ std). Both entropies rise at the morning and evening commute peaks as speed drops.}
    \label{fig:pemsabs}
\end{figure}

\begin{figure}[t]
    \centering
    \includegraphics[width=\linewidth]{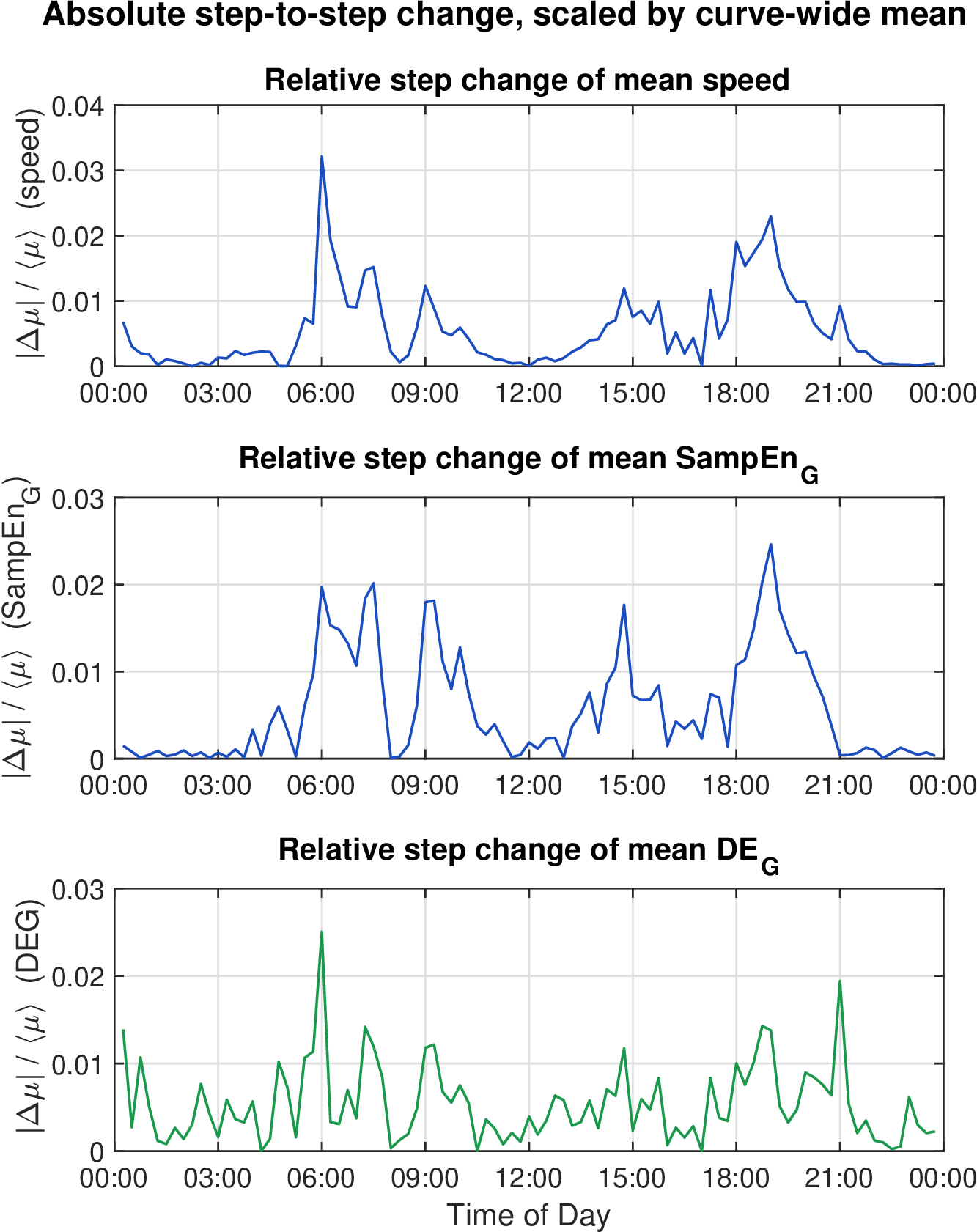}
    \caption{Step-to-step change of each curve in Fig.~\ref{fig:pemsabs}, normalised by its mean. SampEn$_{G}$ peaks align with the speed transitions ($\approx06:00$, $\approx18:30$); DE$_G$ peaks are less synchronised.}
    \label{fig:pemschange}
\end{figure}

The size of the network is associated with persistently congested conditions over the day, which is evident from the mean speed measurements in Fig.~\ref{fig:pemsabs}. We therefore evaluated SampEn$_{G}$ and DE$_G$ over the entire 24-hour interval using a undirected graph with speed as the graph signal, reporting the mean \(\pm\) standard deviation entropy values and their relative step change in mean in Fig.~\ref{fig:pemsabs}, and Fig.~\ref{fig:pemschange}.

Fig.~\ref{fig:pemsabs} reports the 61-weekday profile of mean speed, SampEn$_{G}$, and DE$_{G}$, averaged over $61$ weekdays. Despite the large network size and the persistently congested conditions evident in the mean speed remaining throughout the day, both entropy measures cleanly characterise the two daily congestion peaks corresponding to the morning and evening commutes. SampEn$_{G}$ remains \(\approx1.05\) during the overnight hours and rises to \(\approx1.25\) during the morning (\(06:00–09:00\)) and evening (\(15:00–18:00\)) peaks, mirrored by a decline in mean speed. DE$_{G}$ follows the same qualitative structure, consistent with the free-flow-versus-congestion contrast reported for the FT-AED freeway in Sec.~\ref{Sec:ft-aed}, now reproduced on a network two orders of magnitude larger.

To characterise the transitions, Fig.~\ref{fig:pemschange} plots the absolute step-to-step change of each curve, normalised by their mean. For all three measures the sharpest relative changes occur at the onset and recovery of the rush periods, and the morning transition near \(06:00\) is the most abrupt feature of the day, in which congestion onset is concentrated in a narrow window whereas evening build-up and recovery is more gradual. 

Notably, the SampEn$_{G}$ step-change peaks are tightly aligned with those of mean speed at both the morning (\(\approx06:00\)) and evening (\(\approx18:30\)) transitions, whereas the largest DE$_{G}$ step changes are less synchronised, with a pronounced late-evening peak (\(\approx21:00\)) that lags the corresponding speed transition. This indicates that the SampEn$_{G}$ response is driven by the speed-regime transitions themselves rather than by the absolute congestion level.

Between the two peaks, SampEn$_{G}$ additionally registers moderate step changes through the midday and early-afternoon interval (\(\approx9:00\) and \(\approx14:00–15:00\)), where DE$_{G}$ appears less sensitive to. This suggests that SampEn$_{G}$ flags finer-grained intra-day variability, consistent with the conditional-entropy formulation providing information complementary to Shannon-based measures which the ordinal DE$_G$ measure smooths over.

This analysis was conducted on an undirected graph with speed as the graph signal. We note that the "early-warning" signal from Sec.~\ref{Sec:ft-aed} was not replicated here, and may be attributed to the reasons: (i) the measure is reporting a network-wide summary of a very large-scale traffic network covering nine districts in the California traffic system, where any earlier changes may be diluted; (ii) PeMS was aggregated to 15-minute resolution, in contrast to $30$~s in FT-AED, in which early transition events may be missed or averaged out. Nevertheless, SampEn$_{G}$ retains a clear free-flow-congestion signature over the time scale. The computation over \(18,176\) sensors confirms the practical feasibility of SampEn$_{G}$.

\section{Discussions and Conclusions}\label{Sec:discussions}

In this paper, we introduced SampEn$_{G}$, a graph-based generalisation of SampEn for the analysis of complex dynamical processes on networks. The proposed measure constructs topology-aware patterns by aggregating signals over multi-hop neighbourhoods. In contrast to previous entropy measures grounded in Shannon entropy--PE$_{G}$, DE$_{G}$, and BE$_{G}$--SampEn$_{G}$ is built on correlation integrals with a fixed tolerance $\epsilon$ to quantify the regularity, and reccurrence of local patterns with conditional entropy as their dimension is increased from $m$ to $m+1$. 

We formulated a formal extension of SampEn to graph signals, applicable to directed/undirected, and binary/weighted graphs, and further evaluated on synthetic and real-world datasets systematically. The empirical experiments provide a proof-of-concept that SampEn$_{G}$ unifies 1D time series, 2D images, and arbitrary graph signals within a single framework.

In particular, on directed paths, SampEn$_{G}$ reproduces classical SampEn on the logistic map. On 2D images modelled as signals over regular grids, it yields texture rankings consistent with SampEn$_{2D}$ \cite{sampen_2d}. Our experiment using the Kuramoto model to simulate dynamic organisation over three network topologies demonstrated the effectiveness of SampEn$_{G}$ in: (i) tracking local synchronisation before global synchronisation, discriminating the locally-constrained geometric graphs from globally connected (Barab\'asi-Albert and Erd\H{o}s--R\'enyi networks), (ii) in scenario where distribution of frequency patterns are similar, as in the sine signal from the Kuramoto model, we see that SampEn$_{G}$ was able to continue tracking the network organisation; while DE$_{G}$ fails. This was attributed to the assessment of pattern recurrence on a continuous space, rather than binning the signals and evaluating them as a cumulative frequency distribution.

On real-world networked systems, including weather station, wireless sensor networks, and freeway traffic (FT-AED and PeMS), SampEn$_{G}$ distinguishes between dynamical regimes that are known to differ in variability and organisation (night vs day, free-flow vs congestion). These applications also illustrate that the method is effective for relatively short recordings, as in the Intel Berkeley sensor network ($108$ samples within each window), and adaptable to distinct interpretations of the underlying topology, from geometric distance (weather stations) to wireless communication-reliability.

In the FT-AED freeway experiment, SampEn$_{G}$ computed on the \emph{directed} traffic topology (encoding downstream flow constraints) appears to respond earlier to the onset of congestion than both DE$_G$ and the mean measured speed. We note, however, that this early-warning behaviour was observed in this specific directed-topology setting. In contrast, in the extended PeMS analysis over a \(18k\)-node sensor network, conducted on an \emph{undirected} distance-based similarity graph of state scale, did not show the early-warning advantage. Therefore, the timing advantage should be interpreted as potentially \emph{topology/single freeway-dependent} (and possibly facilitated by directionality), rather than a universal property across traffic networks.

Across the three state-transition case studies on networks---the Kuramoto model (Tables~\ref{Table:KMsd} and~\ref{Table:KMbend}), FT-AED (Fig.~\ref{fig:EntropyFT-AED}), and PeMS (Fig.~\ref{fig:pemschange})---SampEn$_{G}$ shows consistently strong sensitivity to changes in network organisation, often providing information complementary to the Shannon-entropy based DE$_G$. A systematic comparison of directed vs.\ undirected constructions on matched traffic graphs is a natural direction for future work to clarify when directionality yields earlier or sharper transition detection.

At the same time, the investigations with synthetic experiments reveal an important requirement for SampEn$_{G}$: the preservation of local multi-hop neighbourhood distinctiveness. SampEn$_{G}$ relies on matching probabilities defined in respect to a fixed, hard tolerance $\epsilon$, its behaviour therefore differs from those of Shannon-based entropies where the state space is partitioned and each symbol is guaranteed to fall into one of the bins. 

When the topology is very dense and rich in long-range connections, the $L$-neighbourhoods extensively overlap as hop radius expands, with increase in embedding dimension $m$. This is aggravated when the signal structure is dominated by noise, as the averaging operation acts as a low-pass filter, causing the constructed patterns to tend toward a global mean of the graph signal. As a result, most patterns fall within the tolerance $\epsilon$ of each other, and the empirical difference in matching probability between dimension $m$ and $m+1$ becomes insignificant. 


These observations suggest that SampEn$_{G}$ is best suited to systems where (i) the graph has sparse or moderate connectivity with limited long-range connections, so that multi-hop neighbourhoods remain distinguishable, and (ii) the signal dynamics generated attains structure, rather than being completely noise-dominated. In such regimes, short embeddings ($m=1$ or $m=2$) and tolerance with $r\approx0.2$, as in classical SampEn, provide stable interpretable estimates for SampEn$_{G}$. This is consistent with the robustness experiments above in Sec.~\ref{Sec:robustness_construction} and Sec.~\ref{Sec:perturbedge}: SampEn$_G$ remains discriminative on dense graphs as long as connectivity is predominantly local (in Fig.~\ref{fig:sigma1vary}), but it tends to fail when dense connectivity is driven by long-range connections, which destroys difference between multi-hop embeddings which SampEn$_{G}$ can utilise (seen in the edge addition experiment in Fig.~\ref{fig:perturbgap}).

Furthermore, a limitation of SampEn$_G$ is that its absolute values can be comparatively small and are not directly comparable across datasets. This is expected as the graph-based embedding uses walk-weighted multi-hop averages, which can increase the empirical matching probability (especially on graphs with overlapping neighbourhoods). Consequently, SampEn$_G$ is best interpreted as a relative index of irregularity for within-study contrasts computed under identical graph and parameter settings, rather than relying on absolute values or universal thresholds.


Future work includes designing alternative graph-based embeddings or using feature vectors. A limitation is that the speed signal retains detectors reporting zero under the PeMS convention. Masking these speed-blind sensors and recomputing the graph entropies on the reduced network is a natural robustness check for future work. Further research may also extend the framework to other entropy measures, such as multivariate SampEn accounting for different time delays, and transfer entropy for a causality measure on graph signals.

In summary, this paper brings the conditional-entropy based framework into the analysis of dynamics on complex networks by introducing SampEn$_{G}$. While it is not universally optimal, especially in extreme regimes of noise and connectivity, our contribution of SampEn$_G$ offers unique insights and benefits over other existing measures for graph signals. In particular, SampEn enables the assessment of the reccurrence and persistence of patterns within a signal through the threshold $\epsilon$. With appropriate parameter selection, SampEn$_{G}$ offer an invaluable complementary perspective to existing nonlinear measure, supporting the growing trend across fields toward the analysis of high-dimensional, networked dynamical systems.

\section*{Statements and Declaration}

\textbf{Funding} \ This work was supported in part by the Leverhulme Trust via a Research Project Grant (RPG-2020-158). M.-S.M.L. is supported by a PhD studentship of the School of Engineering at the University of Edinburgh, UK. For the purpose of open access, the author has applied a Creative Commons Attribution (CC BY) licence to any Author Accepted Manuscript version arising from this submission.

\medskip
\noindent\textbf{Competing Interests} \
The authors have no relevant financial or non‑financial interests to disclose.


\medskip
\noindent\textbf{Data Availability} No datasets were generated during the
current study. All analysed datasets are publicly available.



%
%


\bibliographystyle{spmpsci}

\bibliography{references_withouturl}


\end{document}